\documentclass[%
 reprint,
 amsmath,amssymb,
 aps,
]{revtex4-2}
\usepackage{graphicx}
\usepackage{dcolumn}
\usepackage{bm}

\begin{document}

\preprint{APS/123-QED}

\title{Geometry controls diffusive target encounters and escape in tubular structures}

\author{Junyeong L. Kim}
\author{Aidan I. Brown}%
 \email{aidan.brown@torontomu.ca}
\affiliation{%
 Department of Physics, Toronto Metropolitan University
}%

\date{\today}

\begin{abstract}
The endoplasmic reticulum (ER) is a network of sheet-like and tubular structures that spans much of a cell and contains molecules undergoing diffusive searches for targets, such as unfolded proteins searching for chaperones and recently-folded proteins searching for export sites. By applying a Brownian dynamics algorithm to simulate molecule diffusion, we describe how ER tube geometry influences whether a searcher will encounter a nearby target or instead diffuse away to a region near to a distinct target, as well as the timescale of successful searches. We find that targets are more likely to be found for longer and narrower tubes, and larger targets, and that search in the tube volume is more sensitive to the search geometry compared to search on the tube surface. Our results suggest ER proteins searching for low-density targets in the membrane and the lumen are very likely to encounter the nearest target before diffusing to the vicinity of another target. Our results have implications for the design of target search simulations and calculations and interpretation of molecular trajectories on the ER network, as well as other organelles with tubular geometry.

\end{abstract}

\keywords{Suggested keywords}
\maketitle

\section{\label{sec:level1}Introduction}

Diffusive motion is essential for transport and function in cell biology~\cite{koch1990diffusion,brangwynne2009intracellular,soh2010reaction,wang2010effects,mogre2020getting}. Intracellular geometry affects the timescale of diffusive search, which can be framed in various ways, including as an effective diffusivity, such as the diffusivity along the mitochondrial axis with cristae present~\cite{dieteren2011solute}; as time evolution of concentrations, such as for calcium moving through the endoplasmic reticulum~\cite{means2006reaction}; as a rate or flux, such as the arrival rate to an organelle surface~\cite{brown2015cluster,brown2017model} or to dispersed receptor regions on an organelle~\cite{berg1977physics}; and as a mean first-passage time, such as the escape time from the the endoplasmic reticulum via channels~\cite{yang2023diffusive} or transit time through nuclear pores~\cite{licata2009first}.

The endoplasmic reticulum (ER) is a cellular organelle with important roles in protein synthesis and transport, protein folding, lipid and steroid synthesis, carbohydrate metabolism, and calcium storage~\cite{schwarz2016endoplasmic}. The geometry of the ER is complex and consists of sheet-like and tubular regions~\cite{westrate2015form}, as well as regions of dense tubular matrices~\cite{nixon2016increased}, forming a single connected network of lumenal volume and membrane surface.
ER tubes are typically 250 -- 750 nm long in yeast cells~\cite{west20113d} and 0.5 -- 5 $\mu$m long in mammalian cells~\cite{schwarz2016endoplasmic,georgiades2017flexibility}, with radii typically ranging from 10 -- 25 nm in yeast cells~\cite{west20113d} and 25 -- 70 nm in mammalian cells~\cite{schroeder2019dynamic}. ER structure can vary with cell cycle stage, cell type, and stress~\cite{schuck2009membrane,schwarz2016endoplasmic}.

A variety of molecules inside the ER undergo diffusive searches for targets. As an important site for protein folding, many unfolded proteins in the ER will search for chaperone proteins in the membrane and lumen to assist folding~\cite{bertolotti2000dynamic,brown2021design}. Once folded inside the ER, many proteins are exported to the Golgi via ER exit sites, which have neck-like entry points off of ER tubes~\cite{weigel2021er,obara2023structural}. Calcium ions can exit the ER through channel proteins in the membrane~\cite{yang2023diffusive}.

A range of theoretical studies have investigated how ER geometry and target distribution affect local diffusive search inside the ER. Pore density, size, and channel length affect calcium escape rates from the ER~\cite{yang2023diffusive}. Slower diffusion through the narrow regions of `pearled' ER tubes is described by the Fick-Jacobs equation~\cite{zwanzig1992diffusion,reguera2001kinetic,mogre2020getting}. Helicoidal ramps between layers of ER sheets facilitate more rapid diffusion between the sheet layers compared to the hypothetical alternative of holes between the sheet layers~\cite{huber2019terasaki}. Various narrow escape problems correspond to molecules undergoing diffusive search for a target inside the ER~\cite{holcman2013control}.

Diffusive search through the broader ER network has also been examined through calculations and simulations. Better-connected ER networks (with more `loops') facilitate faster diffusive search~\cite{brown2020impact}. The effective distribution of targets for rapid search through the ER network depends on the location of sources and the maturation period of searchers, with mobile targets able to more than halve search times~\cite{scott2021diffusive}. The removal of tube connections from experimentally-extracted ER networks accelerates diffusive search if a tube is removed from a dense network region (such as near the nucleus), but slows search if the removed tube connects network regions (more likely in the cell periphery)~\cite{elam2022fast}. ER network tube rearrangement dynamics are sufficiently slow compared to intra-network diffusion that the dynamics have little effect on search times~\cite{scott2023endoplasmic}. Target density in an ER network modulates the effective dimension of the diffusive search process~\cite{mogre2020getting}, and the effective dimension of ER networks controls how search time depends on initial searcher-target separation~\cite{condamin2007first}. Beyond calculations and simulations, recent experimental advances have enabled measurement of the trajectories of individual proteins inside the ER network, providing insight into intra-organelle protein behavior~\cite{obara2024motion}.

Previous calculations and simulations largely modeled search through the ER network as a process occurring along one-dimensional tubes~\cite{brown2020impact,scott2021diffusive,elam2022fast,scott2023endoplasmic}, assuming that reaching the location of a target along the tube axis constituted an encounter with that target. However, a searcher on the ER membrane (surface) or a searcher in the ER lumen (volume) could diffuse past the axial location of the target without encountering the target in two or three dimensions, respectively. In this work we consider how the tube geometry affects the probability that a diffusive searcher will encounter a target once it has reached the axial position of the target, with escape to the axial position of a nearby target as the alternative outcome. We find that along with dependence on tube radius and target size, target density along the tube axis is a key parameter to describe whether a searcher will encounter a target or diffuse to the vicinity of another target. This result suggests that individual sparsely-distributed targets are likely to be encountered once their position is reached along the tube axis, while searchers will often escape from individual densely-spaced targets without an encounter.

\begin{figure}
    \includegraphics[width=\linewidth]{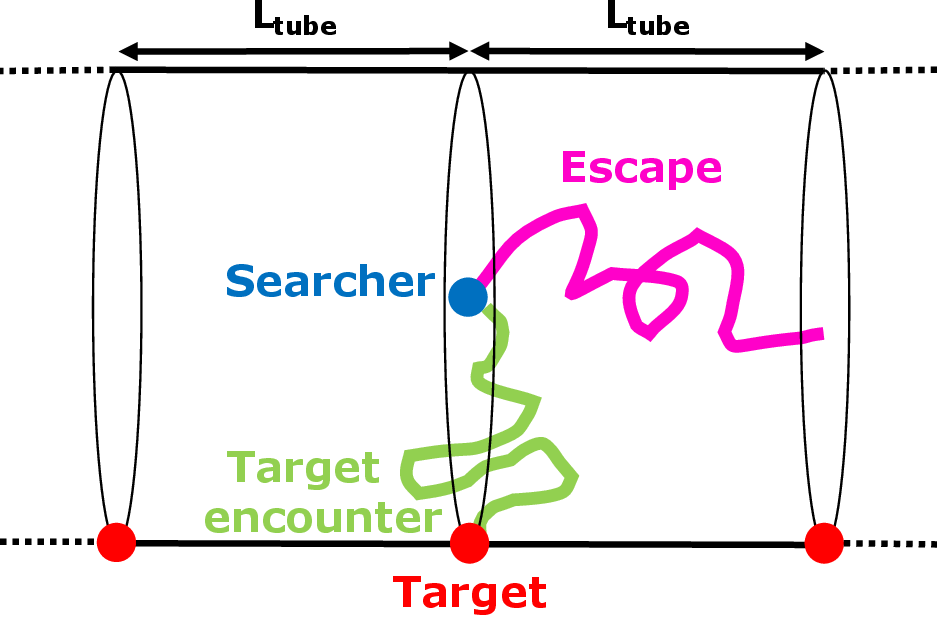}
    \caption{
    Target encounter vs escape in a cylinder. ER tube segment is considered as a cylinder. A searcher molecule (blue dot) begins at the axial position of a target (red dot). Targets (red dots) are separated by a distance $L_{\text{tube}}$ along the tube axis. Undergoing a diffusive trajectory, the searcher reaches one of two outcomes first: encounter the target or escape from the initial target by reaching the axial position of an adjacent target.
    }
    \label{fig:Intro_setup}
\end{figure}

\section{Results}

\subsection{Stationary target encounters}

To investigate diffusive search behavior for a variety of cylindrical tube geometries, we first model the diffusive search for a stationary target on the wall of a tube with two open ends, with searchers either diffusing in the tube volume (representing the ER lumen) or on the tube surface (representing the ER membrane). The stationary target particle with radius $R_{\text{target}}$ is placed on the tube wall at the axial midpoint of the tube a distance $L_{\text{tube}}$ from both ends of the tube, with this axial tube length $L_{\text{tube}}$ representing the distance between targets along the tube axis as shown in Fig.~\ref{fig:Intro_setup}. The tube has radius $R_{\text{tube}}$. The diffusive searcher begins at an axial position offset from the target axial position by $R_{\text{target}}$, with searchers in the tube volume uniformly distributed through the tube cross section and searchers on the tube surface uniformly distributed around the tube circumference. The searcher is considered to have encountered the target once the searcher first arrives within a distance $R_{\text{target}}$ from the target, such that $R_{\text{target}}$ represents the sum of the searcher and target radii. For searchers in the volume this encounter distance $R_{\text{target}}$ is through the three-dimensional tube volume and for searchers on the surface it is along the two-dimensional tube surface. Thus the initial searcher position is at the closest axial position without possible encounter of the target by the searcher. Brownian Dynamics, combined with reflections from the inside tube surface for search in the volume, is used to simulate the diffusive search for the target (see Appendix for details). 

\begin{figure*}
    \centering
    \includegraphics[width=\linewidth]{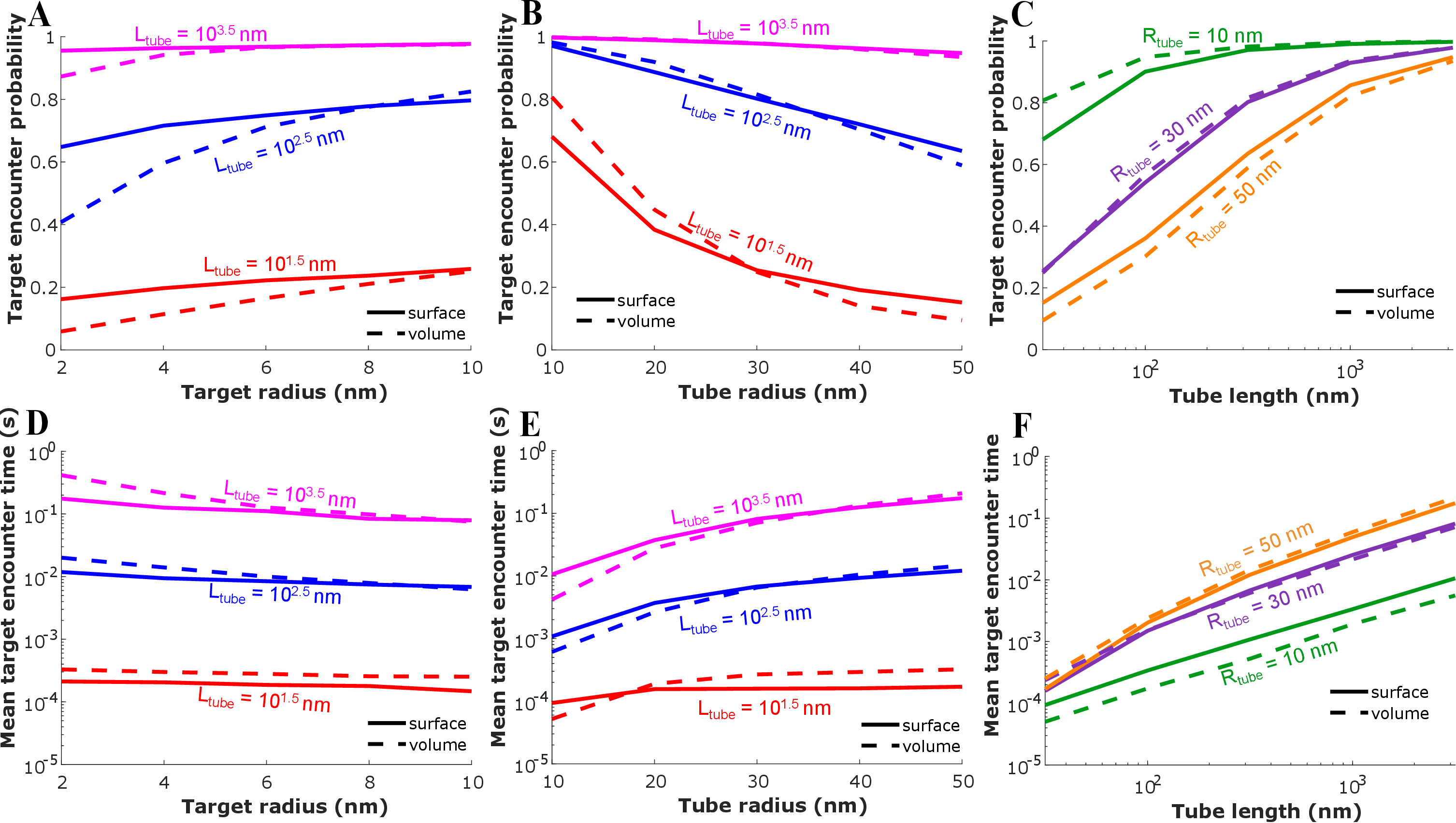}
    \caption{
    Target encounter with a stationary target.
    Target encounter probability for varying (A) target radius, (B) tube radius, and (C) tube length.
    Corresponding mean target encounter time for varying (D) target radius, (E) tube radius, and (F) tube length.
    For all panels, the legend indicates whether the searcher is on the tube surface (solid lines) or in the tube volume (dashed lines). For panels (A), (B), (D), and (E) tube length is indicated by curve color and adjacent corresponding labels. For panels (C) and (F) tube radius is indicated by curve color and adjacent corresponding labels.
    Panels (A) and (D) have tube radius $R_{\text{tube}} = 30 \text{ nm}$ and panels (B), (C), (E), and (F) have target radius $R_{\text{target}} = 10\text{ nm}$. For all panels searcher diffusivity $D = 1\text{ }\mu \text{m}^2/\text{s}$ with $10^4$ samples.
    }
\label{fig:stationary_find}
\end{figure*}

With the tube open at both ends and containing a target, there are two possibilities for each trajectory --- the diffusive searcher will either encounter the target or escape the tube region under consideration by diffusing beyond one of the tube ends --- both outcomes terminate the search. As the tube region under consideration contains one target and both tube ends axially abut the nearest target in each direction (see Fig.~\ref{fig:Intro_setup}), escape represents the close approach of the searcher to another target, ending the encounter opportunity with the target in the axial center of the tube region under consideration.

Figure~\ref{fig:stationary_find} explores how the geometric parameters of this search --- target radius $R_{\text{target}}$, tube radius $R_{\text{tube}}$, and tube length $L_{\text{tube}}$ --- control the probability and mean time to encounter the target. Increasing the target radius $R_{\text{target}}$ increases the target encounter probability (Fig.~\ref{fig:stationary_find}A) and decreases the mean target encounter time (Fig.~\ref{fig:stationary_find}D). Increasing the tube radius $R_{\text{tube}}$ decreases the target encounter probability (Fig.~\ref{fig:stationary_find}B) and increases the mean target encounter time (Fig.~\ref{fig:stationary_find}E). This behavior is expected, as a larger target will be more easily and more quickly encountered and a larger search space will cause a target to become more difficult and slower to encounter.

In Fig.~\ref{fig:stationary_find}A, the encounter probability changes less with increasing target radius for search on the tube surface compared to search in the tube volume, a trend which largely continues for encounter probabilities and times across Figs.~\ref{fig:stationary_find}A,B,D,and E. This behavior aligns with previous work showing that two-dimensional diffusive search depends less on geometric factors than three-dimensional search~\cite{condamin2007first,benichou2010geometry,mogre2020getting}. Variation of target radius (Figs.~\ref{fig:stationary_find}A,D) has a more limited effect on the target encounter probability and target encounter time in comparison to variation of tube radius (Figs.~\ref{fig:stationary_find}B,E).

While increasing the tube length increases the search space, such a tube length increase also increases the target encounter  probability (Fig.~\ref{fig:stationary_find}C). This is in contrast to larger tube radius, which increases the search space and causes a decrease in target encounter probability (Fig.~\ref{fig:stationary_find}B). The increase in encounter probability with tube length occurs because the mean time to escape the tube increases with the tube length, providing a longer time period for the searchers to encounter the target, while varying tube radius does not similarly affect escape time. Similar to increasing tube radius (Fig.~\ref{fig:stationary_find}B), increasing the tube length increases the target encounter time (Fig.~\ref{fig:stationary_find}F). The mean encounter time increases with tube length because the greater mean time period before escape allows longer trajectories to encounter the target before escape.

Figure~\ref{fig:stationary_find}C shows that as the tube length exceeds approximately 1 $\mu$m the target encounter probability approaches one.  Figure~\ref{fig:stationary_find}C shows encounter probabilities for $R_{\text{target}} = 10 \text{ nm}$ so a smaller target will require longer tubes to reach similarly high encounter probabilities. However, the approach of encounter probabilities towards one in Fig.~\ref{fig:stationary_find}C for $L_{\text{tube}}\gtrsim 1\text{ }\mu\text{m}$ suggests that entering the axial proximity of a target along an ER tube confers a high likelihood that the searcher will encounter the target, for targets separated by more than approximately a micrometer. Figures~\ref{fig:stationary_find}D, E, and F indicate the mean target encounter times are $\lesssim 10^{-1}\text{ s}$ for a searcher diffusivity $D = 1\text{ }\mu\text{m}^2/\text{s}$, and would require very long tubes (or corresponding low density targets) for encounter times to reach substantially longer than a second.

In Fig.~\ref{fig:stationary_find}, we use specific ranges for target radius, tube radius, and tube length, and explore similar length scales for these quantities throughout this work.
Protein radii are approximately 1 -- 5 nm for 5 -- 500 kDa proteins~\cite{erickson2009size}. The necks of ER exit sites, which allow entry into the exit site structure, have been observed to be somewhat narrower than 20  -- 30 nm radius ER tubes~\cite{weigel2021er}, suggesting that ER exit site necks are 10 -- 20 nm in radius. Calcium ion pores, enabling transport between the ER lumen and the cytosol, are approximately 1 nm in diameter, but the ER lumen side of the membrane is surrounded by a charged region to facilitate ion capture~\cite{mejia1999unitary,yang2023diffusive}, suggesting a target size of one to several nanometers. The target radius range we explore, from 2 -- 10 nm, covers much of size range of these targets in the ER.

Yeast ER tubes have a typical radius range of 10 -- 25 nm~\cite{west20113d} and mammalian ER tubes have a typical radius range of 25 -- 70 nm~\cite{schroeder2019dynamic}), with cases as low as 14 nm~\cite{tran2021stress}. The tube radius range we explore, from 10 -- 50 nm, covers much of the observed range in ER tube radii.

Tube lengths represent the distance between targets. There are approximately $2\times10^7$ BiP chaperones per mammalian cell~\cite{bakunts2017ratiometric}, largely expected to be confined to the ER, with the overall ER membrane area of $10^4\text{ }\mu\text{m}^2$~\cite{kischuck2023tube} --- this corresponds to (with a typical 50 nm mammalian ER tube radius~\cite{schroeder2019dynamic}) approximately one BiP per 10 nm of tube length. At various levels of unfolded protein stress in the ER~\cite{brown2021design}, unfolded proteins will occupy BiP~\cite{bakunts2017ratiometric,stroberg2018design}, so the expected typical spacing between free BiP is 10 nm or larger. IRE1 proteins are present on the ER membrane with approximately 1/$\mu\text{m}^2$ density~\cite{kischuck2023tube}, which corresponds to a density of approximately one IRE1 every 3 - 10 $\mu\text{m}$ of ER tube length, using typical yeast tube radius of 18 nm~\cite{west20113d} and mammalian tube radius of 50 nm~\cite{schroeder2019dynamic}. Calcium channel densities are estimated to be one per 50 nm or more of tube length~\cite{yang2023diffusive}. ER tubes are typically 250 -- 750 nm long in yeast cells~\cite{west20113d} and 0.5 -- 5 $\mu$m long in mammalian cells~\cite{schwarz2016endoplasmic,georgiades2017flexibility}. The tube length range we explore, from a few tens of nanometers to a few micrometers, covers much of the spacing range for ER targets and the range of ER tube lengths.


\subsection{Escape from a target}

\begin{figure*}
    \includegraphics[width=\linewidth]{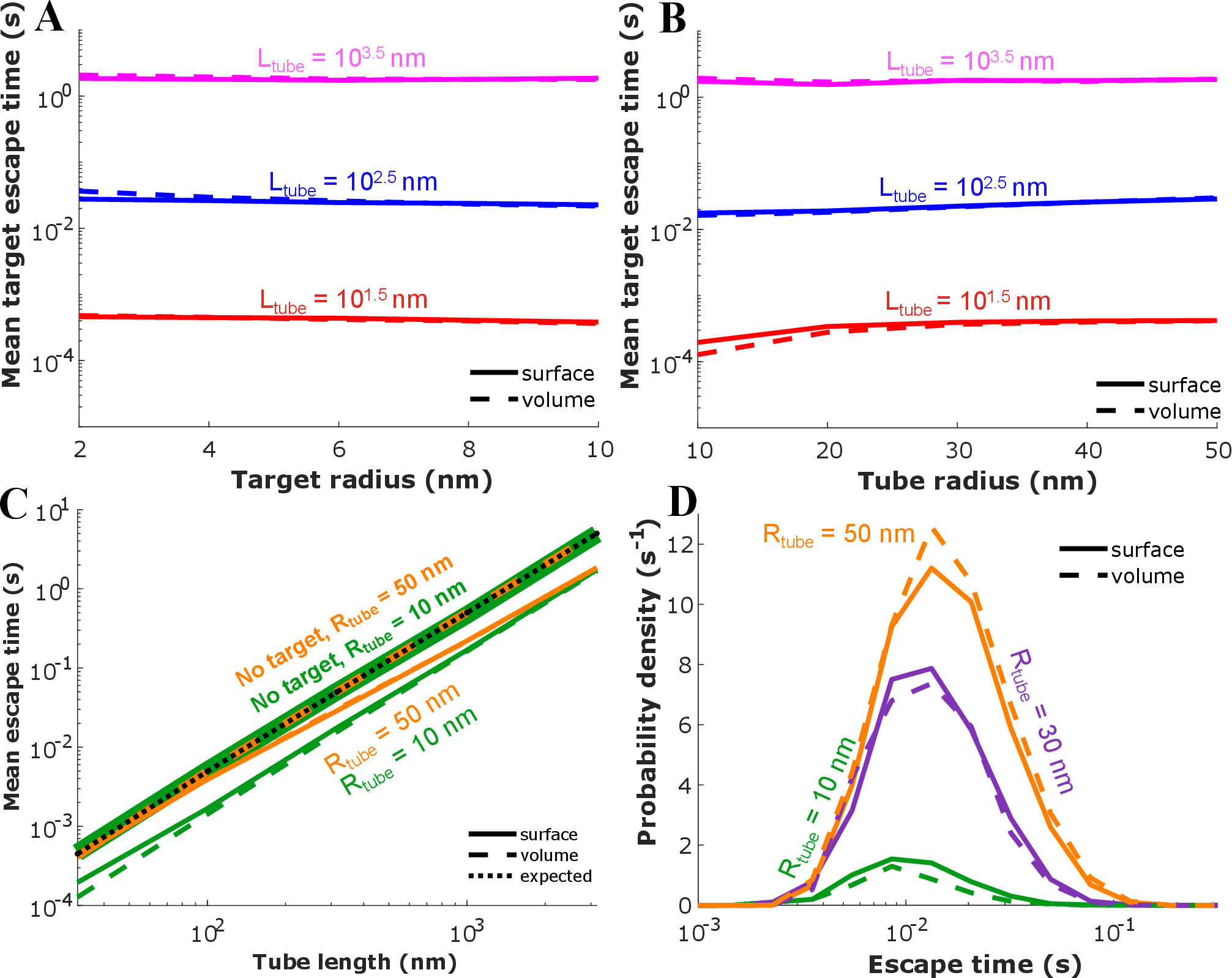}
    \caption{Tube escape with a stationary target.
    Mean escape time with varying (A) target size, (B) tube radius, and (C) tube length.
    (D) shows distribution of escape times for different tube radii with a target radius $R_{\text{target}} = 10\text{ nm}$ and tube length $L_{\text{tube}} = 100\text{ nm}$.
    For all panels, the legend indicates whether the searcher is on the tube surface (solid curves) or in the tube volume (dashed), and the legend of (C) further indicates the expected one-dimensional interval escape from Eq.~\ref{eq:1dsearch}.
    For (A) and (B) tube length is indicated by curve color and adjacent corresponding labels.
    For (C) tube radius is indicated by curve color and whether a target is present by line thickness, as well as with adjacent corresponding labels.   
    For (D) tube radius is indicated by curve color and adjacent corresponding labels.
    (A) uses tube radius $R_{\text{tube}} = 30\text{ nm}$, and (B) and (C) target radius $R_{\text{target}} = 10\text{ nm}$.
    For all panels searcher diffusivity $D = 1\text{ }\mu\text{m}^2/\text{s}$ with $10^4$ samples, except the escape times without targets in (C) are for 2500 samples. 
    }
\label{fig:Stationary_escape}
\end{figure*}

We now examine the trajectories of searchers that escape from the ends of the tube. The escape probability $P_{\text{escape}} = 1 - P_{\text{encounter}}$, with target encounter probability $P_{\text{encounter}}$ examined in Fig.~\ref{fig:stationary_find}A,B,C. Figures~\ref{fig:Stationary_escape}A and B show that target radius and tube radius, respectively, have limited effect on the mean time to escape the tube without encountering the target. While limited, the effect of target radius and tube radius on escape time is not zero, with larger target radius marginally decreasing the mean escape time (Fig.~\ref{fig:Stationary_escape}A) and larger tube radius marginally increasing the mean escape time (Fig.~\ref{fig:Stationary_escape}B). In contrast, Figs.~\ref{fig:Stationary_escape}A, B, and C show that tube length strongly affects the mean escape time, with longer tubes requiring a longer time to escape.

The mean time $\langle t_{\text{1d}}\rangle$ for a particle of diffusivity $D$, initially a distance $x$ from one end of a one-dimensional interval of length $d$,  to first reach one of the two interval ends is~\cite{redner2001guide}
\begin{equation}
    \langle t_{\text{1d}} \rangle = \frac{x(d - x)}{2D} \ .
    \label{eq:1dsearch}
\end{equation}
For the scenario under consideration, the interval length $d = 2L_{\text{tube}}$ and $x = L_{\text{tube}} \pm R_{\text{target}}$. As a searcher in the ER tube is able to escape from the tube ends it is expected that the tube length will largely control the mean time to escape. In the case where there is no target, the searcher will escape with a mean time described by Eq.~\ref{eq:1dsearch} (thick orange and green curves in Fig.~\ref{fig:Stationary_escape}C).

However, with a target present in the tube, the trajectories that escape are those that do not first encounter the target. Figure~\ref{fig:Stationary_escape}C shows that larger tube radii and shorter tube lengths more closely match the mean escape time without a target of Eq.~\ref{eq:1dsearch}. Decreasing the tube radius or increasing the tube length causes the mean escape time to decrease below $\langle t_{\text{1d}}\rangle$ from Eq.~\ref{eq:1dsearch}. Narrower and longer tubes increase the probability that the searcher will encounter the target before escaping (Fig.~\ref{fig:stationary_find}A,B), removing from the ensemble of escape trajectories those trajectories that would have escaped from the tube after a relatively long time period, reducing the mean time to escape. The probability distribution of escape times for various tube radii, shown in Fig.~\ref{fig:Stationary_escape}D, illustrates how target encounter removes escape trajectories as the tube radius decreases, with the probability of longer escape times decreasing more with decreasing tube radius than the probability of shorter escape times.

The escape time from the tube is largely controlled by the tube length. However, increasing target radius or decreasing tube radius increases the target encounter probability and preferentially removes longer escape trajectories, marginally reducing the mean escape time.

\subsection{Mobile targets}

\begin{figure*}
\includegraphics[width=\linewidth]{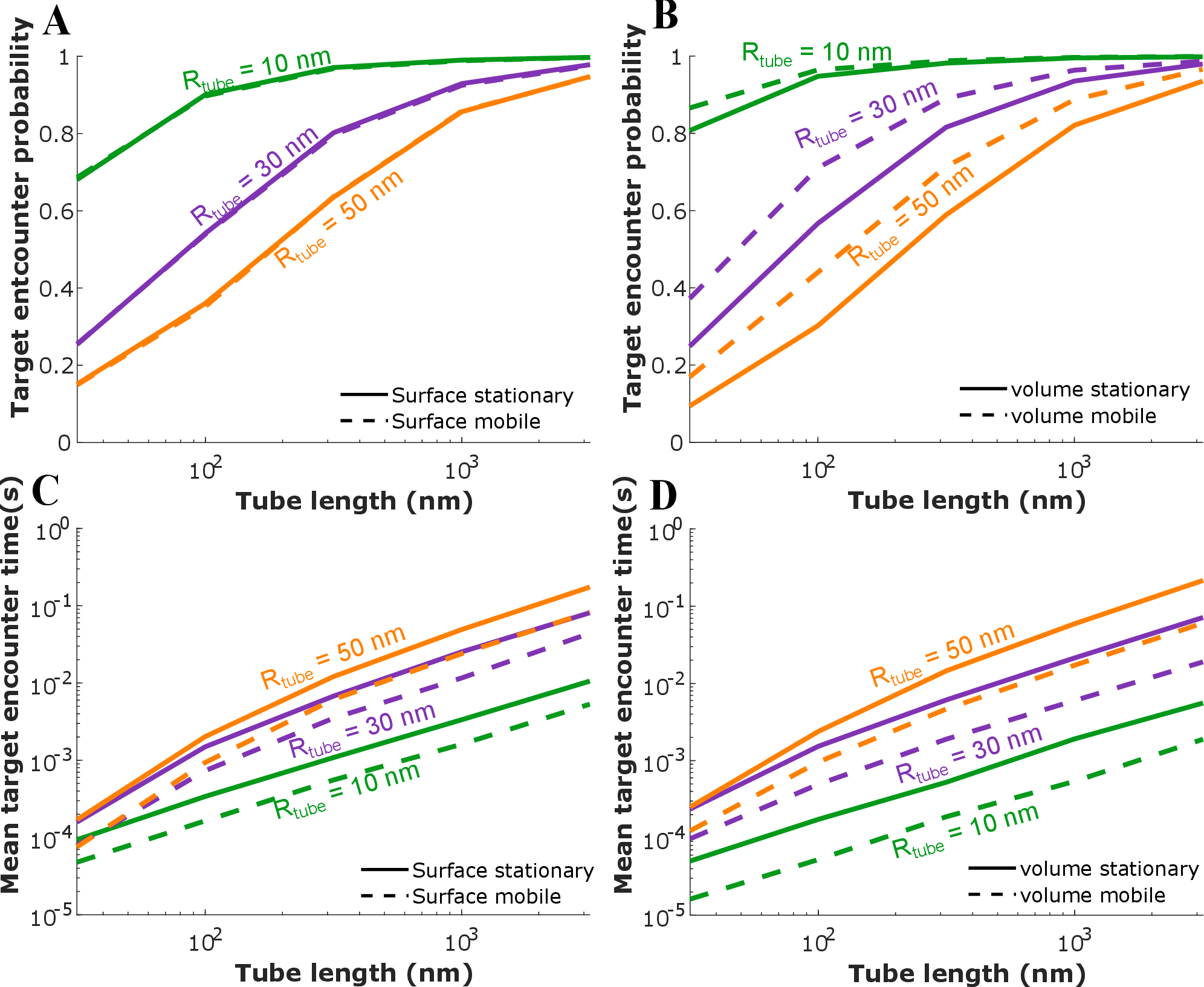}
\caption{
Comparison of stationary vs mobile targets.
Target encounter probability for search (A) on the tube surface and (B) in the tube volume with varying tube length.
Mean target encounter time for search (C) on the tube surface and (D) in the tube volume with varying tube length.
Legends indicate if target is stationary (solid curves) or mobile (dashed).
For all panels, target size $R_{\text{target}} = 10\text{ nm}$, searcher diffusivity $D = 1\text{ }\mu\text{m}^2/\text{s}$ with $10^4$ samples.
}
\label{fig:stationary_vs_mobile_find}
\end{figure*}

Thus far we have examined mobile searchers and stationary targets. We will now explore the scenario where both searcher and target are mobile, effectively turning the target into a searcher as well. The initial condition is similar to search with a stationary target, with both searcher and target separated along the tube axis by a distance $R_{\text{target}}$, and each searcher independently and uniformly distributed through the tube cross section (searchers in the volume) or around the tube circumference (searchers on the surface). The encounter condition is unchanged, such that the two searchers are considered to have encountered one another if they are within a distance $R_{\text{target}}$.  This initial condition represents the closest the two searchers can approach along the tube axis without a possible encounter. Rather than fixed axial positions beyond which searchers are considered to have escaped, the escape condition is that the two searchers have become separated along the tube axis by the tube length $L_{\text{tube}}$, equivalent to the axial distance to the escape threshold with a stationary target.

Target encounter probabilities and mean times are compared for stationary and mobile targets in Fig.~\ref{fig:stationary_vs_mobile_find}. For search on the tube surface, the target encounter probability for mobile targets is unchanged compared to stationary targets (Fig.~\ref{fig:stationary_vs_mobile_find}A). However, for search in the volume, mobile targets have an increased the target encounter probability relative to stationary targets (Fig.~\ref{fig:stationary_vs_mobile_find}B). Although the target encounter probability for search on the tube surface is unchanged between stationary and mobile targets (Fig.~\ref{fig:stationary_vs_mobile_find}A), the mean time to encounter the target decreases for mobile targets relative to stationary targets for search on the tube surface (Fig.~\ref{fig:stationary_vs_mobile_find}C).  For search in the tube volume, mobile targets also exhibit shorter target encounter times than stationary targets (Fig.~\ref{fig:stationary_vs_mobile_find}D).

For searchers and targets on the surface, changing from a stationary to a mobile target affects the encounter time but not the encounter probability. Whether mobile or stationary, a surface target covers the same surface area and fraction of the tube circumference. Two mobile searchers on the surface can instead be considered as a fixed target and a searcher with twice its initial diffusivity, without loss of generality. In this case, the encounter probability will be unchanged from the initial scenario with a mobile searcher and a stationary target on the tube surface (Fig.~\ref{fig:stationary_vs_mobile_find}A), but the target encounter time will be halved (Fig.~\ref{fig:stationary_vs_mobile_find}C).

In contrast to search on the surface, changing from stationary to mobile targets for search on the volume affects both the encounter probability and encounter time (Fig.~\ref{fig:stationary_vs_mobile_find}B,D). Stationary volume targets were positioned on the tube wall (with fixation to the tube wall enabling the target to remain stationary). Mobile volume targets can explore the tube volume, effectively increasing the size of the target by increasing the available volume within $R_{\text{target}}$ of the target, causing the target encounter probability to be higher for mobile targets than stationary targets (Fig.~\ref{fig:stationary_vs_mobile_find}B). Similar to search on the surface, an effectively doubled diffusivity of the relative searcher positions for a mobile target compared to a stationary target leads to a reduction in the target encounter time for searchers in the volume (Fig.~\ref{fig:stationary_vs_mobile_find}D).
This is a similar to previous target search results on one-dimensional models of ER networks, where the mobile targets reduce their encounter time by more than half compared to stationary targets~\cite{scott2021diffusive}.

\subsection{Comparing searcher and targets in the tube volume and on the tube surface}

\begin{figure*}
    \includegraphics[width=\linewidth]{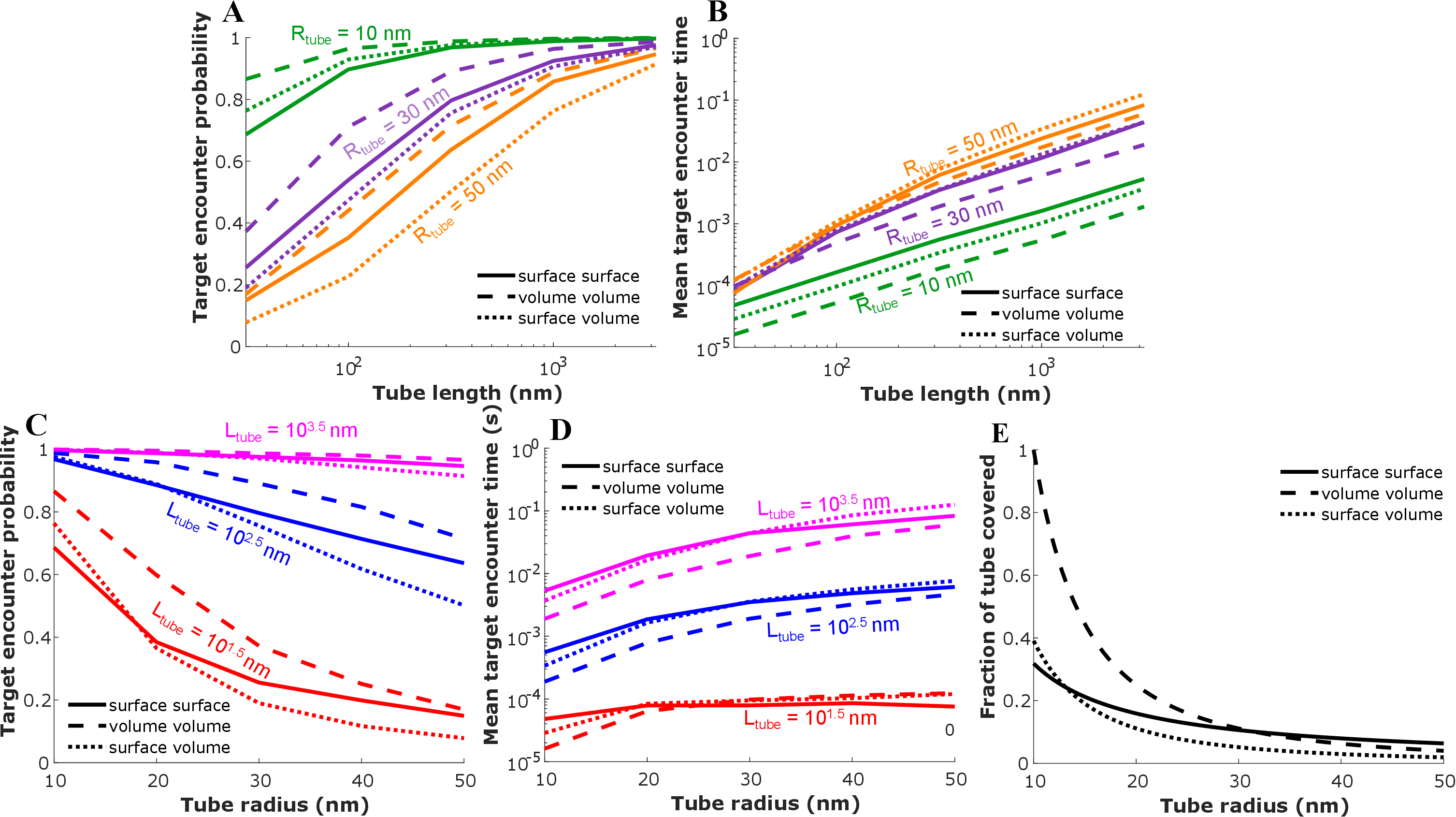}
    \caption{Comparison of searchers both on tube surface, in tube volume, or one on the surface and one in the volume.
    (A) Target encounter probability and (B) mean target encounter time as tube length is varied. Line color and adjacent corresponding labels indicate tube length.
    (C) Target encounter probability and (D) mean target encounter time as tube radius is varied. Line color and adjacent corresponding labels indicate tube radius.
    For panels (A) - (D) target size $R_{\text{target}} = 10\text{ nm}$ and diffusivity $D = 1\text{ }\mu\text{m}^2/\text{s}$, with $10^4$ samples.
    (E) shows fraction of tube covered by a target with $R_{\text{target}} = 10\text{ nm}$. For search entirely on the surface, this is the fraction of the circumference covered by a target. For search entirely in the volume, this is the fraction of the tube cross section covered by a target in the volume that does not overlap the tube circumference. For one searcher on the surface and one searcher in the volume, this is the fraction of the tube cross section covered by a target on the tube surface.
    For all panels legend indicates whether searchers are both on the surface (solid curves), both in the volume (dashed), or one searcher is in the volume and the other is on the surface (dotted).
    }
    \label{fig:3_type_compare}
\end{figure*}

We have examined search with mobile searchers either both in the tube lumen volume or both on the membrane surface. We will now directly compare these two types of searches as well as a search with one searcher in the tube lumen and the other on the membrane surface. Figure~\ref{fig:3_type_compare}A shows that the target encounter probability increases with tube length across search types, as for earlier searches. For all tube radii the highest encounter probability is for search occurring entirely in the lumen volume. For large tube radii the next-to-highest encounter probability is for search occurring entirely on the membrane surface, and the lowest encounter probability is for one searcher in the tube lumen and the other on the membrane surface. However, for smaller tube radii, the encounter probability for search entirely on the membrane surface is lower than for one searcher in the volume and the other on the surface. A corresponding pattern (low encounter probability corresponding to longer encounter time) is seen in Fig.~\ref{fig:3_type_compare}B, with larger tube radii exhibiting the longest encounter times for one searcher in the tube lumen and the next-longest encounter times for search on the membrane surface, and smaller tube radii with longest encounter times for both searchers on the surface and next-to-longest encounter times for one searcher in the volume and the other on the surface. Across radii, the shortest encounter times are for both searchers in the volume. 

The change in the relative encounter probabilities and times between search entirely on the surface and search with one searcher on the surface and the other in the volume for different tube radii is examined in Figs.~\ref{fig:3_type_compare}C and D. Figure~\ref{fig:3_type_compare}C shows that the target encounter probability changes more with tube radius for one searcher each in both the volume and surface compared to both searchers on the surface, allowing the encounter probability for one searcher each in the volume and on the surface to change from a higher probability than both searchers on the surface for small tube radius to a lower probability at large tube radius. There is a similar pattern for encounter time (Fig.~\ref{fig:3_type_compare}D), as one searcher on the surface and one on the volume changes from a shorter encounter time than both searchers on the surface at small radius to a longer encounter time for larger tube radii.


Adjusting the tube radius leads to more substantial changes in encounter probability and time for one searcher each in the volume and on the surface compared to both searchers on the surface. This is consistent with our earlier results where tube geometry has less impact on search on the tube surface compared to search in the tube volume. A change in the relative encounter probabilities and times with tube size for these two search types aligns with the characteristics of these two search types. Encounters for searchers entirely on the surface are mediated by distance along the tube surface, while encounters for one searcher in the volume and one searcher on the membrane are mediated by distance through the volume. For wide tubes, the fraction of the circumference covered by a target (for search entirely on the surface) is larger than the fraction of the cross section covered by the target (for one searcher on the surface and one searcher in volume). As the tube becomes narrow, this order switches, and more cross section is covered compared to circumference (Fig.~\ref{fig:3_type_compare}E). Thus, as the tube narrows, there is less cross-section available for a searcher in the volume to avoid its search partner on the surface, compared to both searchers on the surface.

\subsection{Finite reaction rate}

\begin{figure*}
    \includegraphics[width=\linewidth]{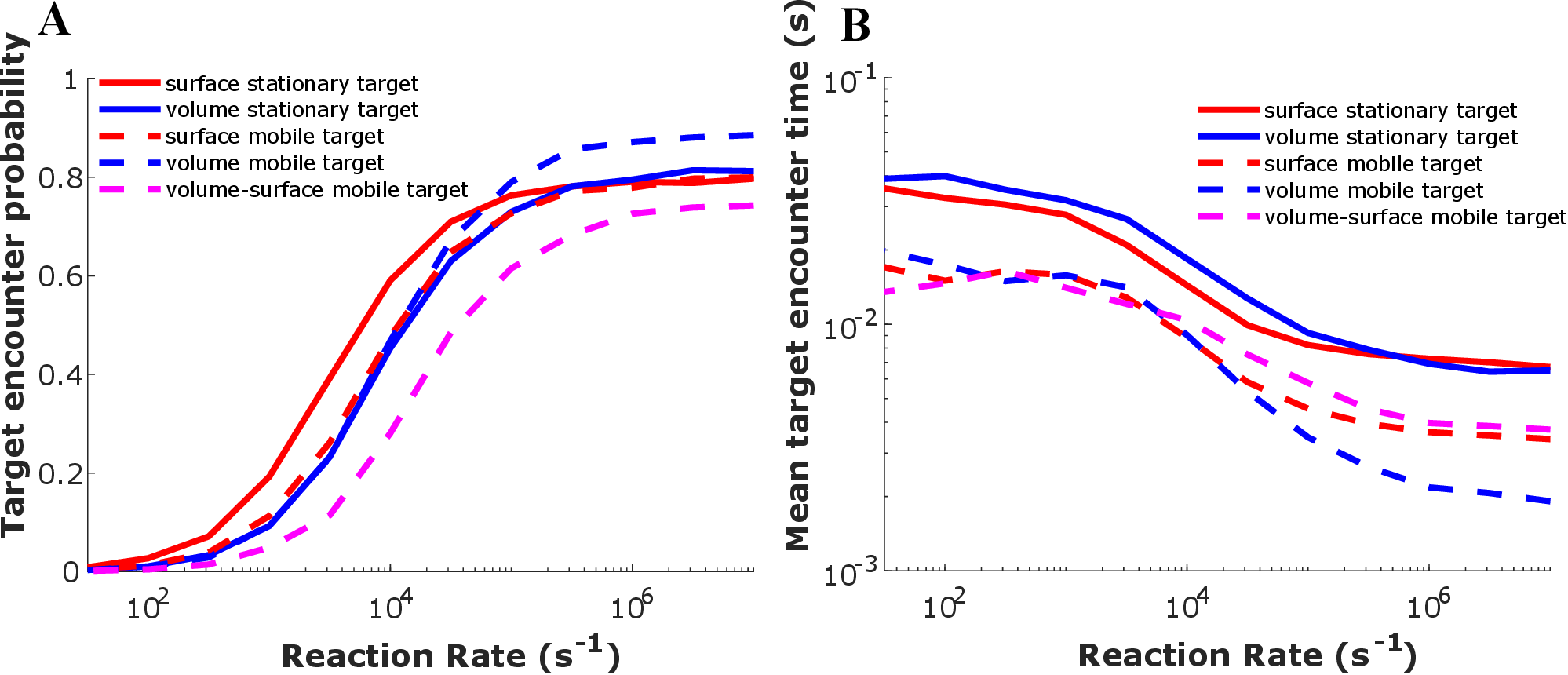}
    \caption{Finite reaction rates.
    (A) Target encounter probability and (B) mean target encounter time as the searcher reaction rate is varied. Legend indicates type of search (on surface, in volume, or one searcher in each; and whether target is stationary or mobile). Tube radius $R_{\text{tube}}= 30 \text{ nm}$, target radius $R_{\text{target}} = 10 \text{ nm}$, tube length $L_{\text{tube}} = 10^{2.5}$ nm, and diffusivity of searcher and mobile targets $D = 1\text{ }\mu\text{m}^2/\text{s}$. Most data points are averaged over $10^4$ trajectories with $2\times10^4$ samples for reaction rates of $10^{1.5}\text{ s}^{-1}$ and $10^{2}\text{ s}^{-1}$ due to high escape probability.}
    \label{fig:Reaction_rate}
\end{figure*}

Thus far we have considered the case of diffusion-limited encounters, corresponding to reactions with infinite reaction rate that undergo instant encounter of a searcher with a target once the searcher is within a distance $R_{\text{target}}$. However, the reaction rate may not be effectively infinite~\cite{collins1949diffusion,hanggi1990reaction}, and could depend on alignment of protein domains, availability of a pore, or other factors~\cite{berg1985diffusion}. We now consider the case where encounters occur with a finite reaction rate once the searcher is within a distance $R_{\text{target}}$ of the target. 

Figure~\ref{fig:Reaction_rate}A shows that for sufficiently low reaction rates ($\lesssim 10^{2}\text{ s}^{-1}$) the target encounter probability is near zero. The target encounter probability increases with the reaction rate until plateauing above a reaction rate of $10^{5} \text{ s}^{-1}$ -- $10^6 \text{ s}^{-1}$. This change in the target encounter probability as the reaction rate increases suggests that sufficiently low reaction rates lead to minimal target encounter and sufficiently high reaction rates nearly guarantee a reaction once the target diffuses within a distance $R_{\text{target}}$.

For a diffusive process the mean squared displacement in $n$ dimensions is $\langle r^2\rangle = 2nDt$, where $t$ is time. Rearranging relates time to mean squared displacement $\langle r^2\rangle$ of $t = \langle r^2\rangle/(2nD)$. With a displacement representing the target reaction range $R_{\text{target}} = 10\text{ nm}$ and $D = 1\text{ }\mu\text{m}^2/\text{s}$, this time is $t = 5\times10^{-5}\text{ s}/n$, which is $2.5\times10^{-5}\text{ s}$ for $n=2$ surface search and $1.67\times10^{-5}\text{ s}$ for $n=3$ volume search. While this estimate is not a precise calculation of the time spent by a diffusive trajectory within a distance $r$ of a target, it does provide an order of magnitude-level estimate of the time spent by the searcher near the target. This dwell time estimate predicts reaction rates of approximately $4\times10^4\text{ s}^{-1}$ for surface search and $6\times10^4\text{ s}^{-1}$ for volume search for the beginning of saturation in encounter probability, which approximately aligns with the reaction rates $10^{5} \text{ s}^{-1}$ -- $10^6 \text{ s}^{-1}$ at which the encounter probability plateaus in Fig.~\ref{fig:Reaction_rate}A.

As the target encounter probability increases from nearly zero to approaching a plateau as the reaction rate increases over the range shown in Fig.~\ref{fig:Reaction_rate}A, the mean target encounter time in Fig.~\ref{fig:Reaction_rate}B decreases by less than an order of magnitude. The searcher explores the tube until the searcher escapes from the tube end or reacts with the target. At high reaction rates and for stationary targets in Fig.~\ref{fig:Reaction_rate}B, the encounter times agree with the encounter times for infinite reaction rate from Fig.~\ref{fig:stationary_find}F. At lower reaction rates, particularly those below $10^{5} \text{ s}^{-1}$ -- $10^6 \text{ s}^{-1}$ above which the encounter probability plateaus, the target encounter time modestly increases above the encounter time for infinite reaction rate. Across reaction rates, the finite-reaction rate search-and-reaction process is competing with escape. We thus expect the mean target encounter time to not significantly exceed the mean escape time even for quite low reaction rates, but instead for the target encounter probability to decrease. The diffusion-limited escape time from Fig.~\ref{fig:Stationary_escape}B corresponding to parameters and stationary target case of Fig.~\ref{fig:Reaction_rate} is approximately $2\times10^{-2}\text{ s}$, such that the low reaction rates of Fig.~\ref{fig:Reaction_rate}B modestly increase from the encounter time for the infinite reaction rate.

\subsection{Target encounter without escape}

\begin{figure*}
    \includegraphics[width=\linewidth]{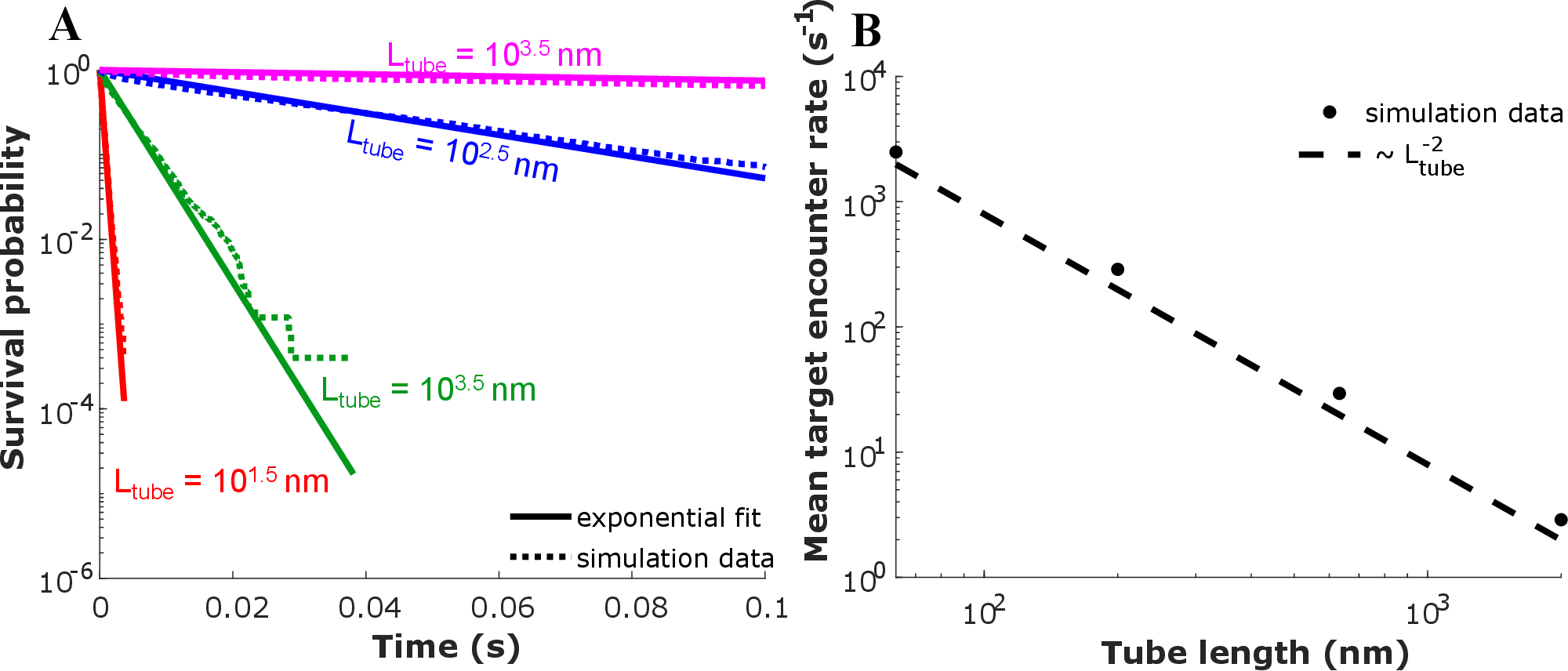}
    \caption{Target encounter without tube escape.
    (A) Survival probability (probability that a searcher has not encountered the target) vs time for a stationary target on the tube surface.
    Curve labels indicate tube length.
    Noting logarithmic axis for survival probability $S(t)$, straight lines are exponential fits $S(t) = e^{-kt}$, with $k$ the fitted encounter rate.
    (B) Black points are the fitted encounter rates $k$ from (A). Dashed curve is inverse tube length squared $1/L_{\text{tube}}^2$ scaling.
    For both panels target radius $R_{\text{target}} = 10\text{ nm}$, tube radius $R_{\text{tube}} = 10\text{ nm}$, and searcher diffusivity $D = 1\text{ }\mu\text{m}/\text{s}$ with 2500 sampled encounters.}
    \label{fig:cumulative_rate}
\end{figure*}

Thus far we have examined a scenario where a searcher either encounters a target inside a tube or escapes from the tube. We now briefly explore the rate at which a searcher encounters a target when escape is not permitted (i.e., tube ends are reflecting). An equivalent scenario is that reaction with any target is considered rather than one specific target, and if a searcher diffuses sufficiently far along the tube axis from one target, it will pass another target rather than escape. We examine a stationary target on the tube surface with a searcher beginning at a uniformly-distributed random position in the tube beyond a distance $R_{\text{target}}$ from the target. Fig.~\ref{fig:cumulative_rate}A shows the time evolution of the survival probability, or the probability that a searcher has not yet found a target. The survival probability appears approximately exponential, $P_{\text{survival}}(t) \simeq \exp(-k t)$, with target encounter rate $k$. Figure~\ref{fig:cumulative_rate}A shows that the target encounter rate $k$ decreases with tube length, with Fig.~\ref{fig:cumulative_rate}B showing the rate dependence on tube length is consistent with $k\sim L_{\text{tube}}^{-2}$. This dependence is similar to other diffusive search timescales that depend on the square of a lengthscale~\cite{berg1977physics,redner2001guide,grebenkov2018strong}.

\section{Discussion}


Proteins, ions, and other molecules diffuse in the endoplasmic reticulum (ER) searching for targets such as protein binding partners and exit sites. Much of the ER is composed of a network of connected tubes that extend through the cell volume. Using Brownian dynamics simulations, we explore how the geometry of diffusive search in a tube, particularly the distance between potential targets along the tube, affects the success and time period for the target search process.

Previous work on diffusive search in the ER network has focused on the organelle-scale geometry of the ER, considering how connectivity via `loops'~\cite{brown2020impact}, searcher maturation and source and target distribution~\cite{mogre2020getting,scott2021diffusive}, individual network edges~\cite{elam2022fast}, and network rearrangement dynamics~\cite{scott2023endoplasmic} affect diffusive target search; or how local tube geometry affects target search via target density and tube geometry~\cite{yang2023diffusive}, how tube radius changes along the tube axis affect diffusivity~\cite{zwanzig1992diffusion,reguera2001kinetic,mogre2020getting}, and control of mean first-passage times in cylindrical geometries~\cite{holcman2013control}. Our work focuses on the local geometry while connecting to larger scales by allowing searchers to escape the tube region under consideration without a target encounter.

We consider a scenario where a searcher that has nearly reached the axial position of a target in a tube either encounters the target or escapes from the target by reaching one of the two tube ends. As expected, larger targets and narrower tubes lead to a higher probability of finding the target and a shorter mean search time. Longer tubes, despite expanding the tube region in which to search, increase both the probability of finding the target and the mean search time because escape time increases with tube length, providing a longer time period for the searcher to find the target. Tube length more strongly affects encounter probability and time in comparison to tube radius and target sizes, when varied across a physiological range.

This finding has implications for simulations and calculations that treat the ER network as one-dimensional, as well as for experimental approaches that measure the searcher position along the length of ER network tubes. For targets that are sparsely distributed along the length of ER tubes, a specific target is likely to be encountered on the tube surface or in the volume if the target axial position along the tube is reached by the searcher. Conversely, for targets that are densely spaced along the length of ER tubes, a specific target is unlikely to be encountered following arrival at the target axial position, as the searcher is likely to come into close axial proximity of another target before encountering the first target. Thus simulations and experimental measurements that assume target encounter once the axial target position is reached are more accurate for widely-spaced targets, and less accurate for densely-spaced targets. For yeast cells (tubes with length 250 -- 750 nm and radius 10 -- 25 nm~\cite{west20113d}) spacing between targets of a few hundred nanometers, corresponding to a single target per tube, provides a probability of target encounter exceeding approximately $75\%$. For mammalian cells (tubes with length 0.5 -- 5 $\mu$m and radius 25 -- 70 nm~\cite{west20113d}) spacing between targets of $\gtrsim 1\text{ }\mu\text{m}$, corresponding to one to a few targets per tube,  provides a probability of target encounter exceeding approximately $75\%$.

We found that the timescale to encounter a target in three dimensions once the target position is reached along the tube axis is less than one second for a diffusivity $D = 1\text{ }\mu\text{m}^2/\text{s}$. Previous calculations suggested that the timescale to diffusively find a particular tube axis position in a mammalian ER network is hundreds of seconds or more for a searcher with diffusivity $D = 1\text{ }\mu\text{m}^2/\text{s}$~\cite{brown2020impact}. The time period until target encounter following arrival at the target axial position in the ER network is likely more than an order of magnitude shorter than the search for the axial position in the network. This suggests that if target encounter occurs, the time of target encounter will closely follow arrival at the axial position of the target in a simulation or during an experimental measurement.

As targets become more dense along the tube length, a searcher starting near one target is increasingly likely to diffuse into the proximity of another target rather than encounter the initial target. This is an example of diminishing returns in diffusive target capture, similar to limited increase in flux through higher density calcium channels~\cite{yang2023diffusive} and absorbing patches on cell surfaces~\cite{berg1977physics}.

The probability of escape from the tube ends, as the alternative to target encounter, decreases with tube length and target radius and increases with tube radius. We find that target and tube radii have limited effect on escape time. Mean escape time is largely controlled by tube length, similar to escape on a one-dimensional interval~\cite{redner2001guide}. Encounter times cannot become too much slower than escape times, as they represent competing processes, such that mean encounter times will be similar to mean escape times. While wider tubes and smaller targets closely follow the predicted mean escape time for a one-dimensional interval, narrower tubes and larger targets decrease the mean escape time. This is because as the tube is narrowed, longer escape trajectories are preferentially converted to target encounter trajectories.

Recent work explores similar diffusive search problems in cylindrical geometry similar to our investigation.
Grebenkov and Skvortsov~\cite{grebenkov2022mean} calculate mean first-passage times to small targets in elongated domains, but do not consider the possibility of escape.
Holcman and Schuss~\cite{holcman2013control} consider the mean first-passage time for target encounter vs escape in a cylindrical geometry for searchers beginning at specific positions. Compared to these previous studies, we have further considered the possibility of escape, a distributed initial searcher position, targets in the tube volume, and mobile targets.

We explore search occurring on the tube surface, in the tube volume, and both on the surface and in the volume. We find that search in the volume is more sensitive to the tube and target geometry, consistent with previous work showing that diffusive search in two dimensions depends less on search geometry than search in three dimensions~\cite{condamin2007first,benichou2010geometry,mogre2020getting}. Across the physiological parameter ranges we have explored, we find that search in the volume is typically (but not always) more likely and more quickly to lead to target encounter. For search on the surface, changing from stationary to mobile targets leaves encounter probabilities unchanged but decreases encounter times by a factor of two, while changing from stationary to mobile targets for search in the volume both increases the target encounter probability and decreases the encounter times. For equivalent parameters, having both searchers in the volume is more likely to lead to target encounter and has more rapid target encounter than having both searchers on the surface, which in turn has more likely and more rapid target encounter than one searcher in the volume and the other on the surface. Small tube radii is an exception, reversing the relative encounter probabilities and times between both searchers on the surface and one each in the volume and on the surface.

Much of our results assume infinite reaction rate, such that any contact between searcher and target leads to a target encounter. We find that adjusting a finite reaction rate causes encounter probabilities that approach zero for reaction rates $\lesssim 10^{2}/\text{s}$ with searcher diffusivity of 1 $\mu$m$^2$/s and encounter probabilities that plateau at nearly the infinite-reaction-rate probability for $\gtrsim 10^6/\text{s}$. While the encounter probability substantially changes with reaction rate, the mean target encounter time is less affected over a wide reaction rate range, due to competition between between target encounter and tube escape. Thus while reaction rate will significantly impact whether a target is found, the reaction rate will only have a limited effect on the typical time for successful searchers to find a target before diffusing to the proximity of another target.

Most of our results consider the scenario where a searcher may either encounter a target or escape from the tube ends. We also found the rate at which the searcher would find a target when confined to a tube region (i.e., the searcher cannot escape the tube) has scaling consistent with the inverse square of tube length, similar to scaling found in previous work for diffusive search in confined geometry~\cite{berg1977physics,redner2001guide,grebenkov2018strong}. Exponential dependence of particle survival with time for a wide range of tube lengths suggests that even for widely-spaced targets, the encounter of a searcher with a target may be considered as a Poisson process (with geometry-dependent rate) for appropriate initial conditions and target spacing distribution.

Using Brownian dynamics simulations in a cylindrical geometry representing ER tubes, we have described how tube geometry and search type affect the probability of a diffusive searcher to find a target and the time to find the target, notably showing that searchers are more likely to find a target vs escape in longer tubes, which represent widely-spaced targets. These findings have implications for designing simulations and interpreting experimental measurements.

\begin{acknowledgments}
This work was supported by a Natural Sciences and
Engineering Research Council of Canada (NSERC) Discovery Grant (A.I.B.) and by start-up funds provided by the Toronto Metropolitan University Faculty of Science (A.I.B.), and was enabled
by computational resources provided by the Digital Research Alliance of Canada (alliancecan.ca), including a cluster usage resource allocation (A.I.B.).
The authors thank Sean Cornelius and Eric De Giuli (Toronto Metropolitan University) for useful discussions and feedback.
\end{acknowledgments}

\appendix

\section{Further method details}


\begin{figure}
    \includegraphics[width=0.9\linewidth]{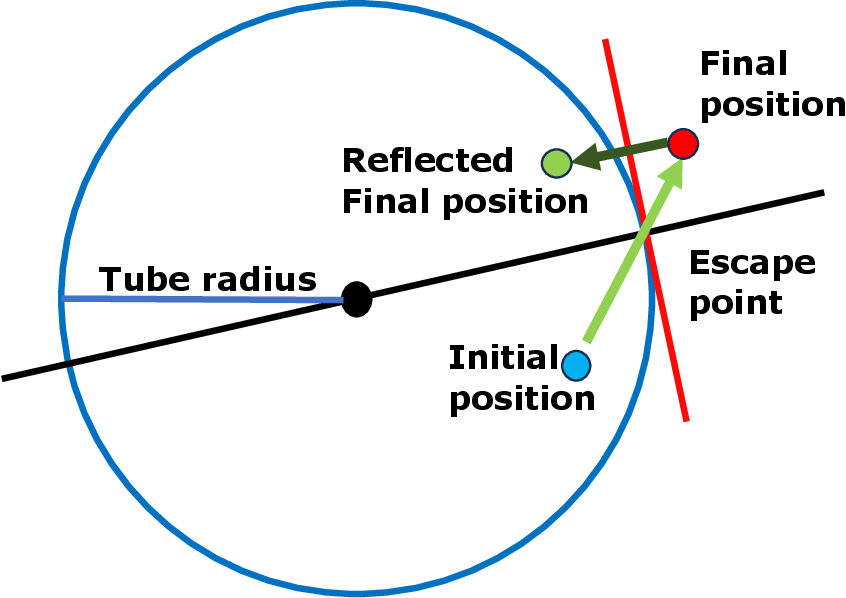}
    \caption{Particle reflection. When a three-dimensional trajectory exits the tube, it is reflected back into the tube volume. The particle, initially at the blue dot, takes a step (yellow arrow) to a final point (red dot) outside the tube. A tangent line (red line) is found at the point at which the trajectory exited the tube (black circle) and the final point outside the circle (red dot) is moved to perpendicular to the tangent line to a reflected position (green dot) equidistant to but on the other side of the tangent line.}
    \label{fig:bounce}
\end{figure}

Stochastic particle trajectories are generated with Brownian dynamics: each timestep $\Delta t$ the particle moves a distance randomly sampled from the absolute value of a normal distribution of mean zero and variance $\sigma^2 = 4D\Delta t$ for diffusion on the tube surface and $\sigma^2 = 6D\Delta t$ in the tube volume, and direction also randomly and uniformly sampled, with $\Delta t = 10^{-2} \text{ s}$. Diffusion on the tube surface is two dimensional with periodic boundary conditions in one direction, and in the tube volume diffusion is three dimensional. We model the ER tube as a cylinder. For three-dimensional diffusion in the tube volume, if a Brownian dynamics step yields a new particle position outside of the cylinder, the trajectory vector from the initial particle position inside the cylinder to the new particle position outside of the cylinder is reflected from the cylinder wall to remain inside the cylinder. Reflection uses a `midpoint method' where the point the trajectory exits the tube cross-section is found, the line tangent to the circle at that exit point defined, a line perpendicular to the tangent and through the (forbidden) trajectory point defined, and the reflected trajectory point inside the tube being on this perpendicular line the same distance from the tangent line as the forbidden point, but on the opposite side of the tangent line (see Fig.~\ref{fig:bounce}).

For Figures~\ref{fig:stationary_find} to \ref{fig:Reaction_rate}, particles begin each simulation at an axial position at the edge of the target. For two-dimensional search entirely on the surface, one particle is set at an angular position defined as zero, while the initial angular position of the other particle is selected randomly and uniformly around the tube circumference. For three-dimensional search entirely in the volume with a stationary target, the mobile searcher randomly and uniformly samples from the tube cross-section. For three-dimensional search entirely in the volume with a mobile target, both the searchers randomly and uniformly sample from the tube cross-section. For one searcher on the tube surface and the other in the tube volume, the searcher on the surface begins at an angular position defined as zero and the searcher in the volume randomly and uniformly samples from the tube cross-section.


\begin{thebibliography}{46}%
	\makeatletter
	\providecommand \@ifxundefined [1]{%
		\@ifx{#1\undefined}
	}%
	\providecommand \@ifnum [1]{%
		\ifnum #1\expandafter \@firstoftwo
		\else \expandafter \@secondoftwo
		\fi
	}%
	\providecommand \@ifx [1]{%
		\ifx #1\expandafter \@firstoftwo
		\else \expandafter \@secondoftwo
		\fi
	}%
	\providecommand \natexlab [1]{#1}%
	\providecommand \enquote  [1]{``#1''}%
	\providecommand \bibnamefont  [1]{#1}%
	\providecommand \bibfnamefont [1]{#1}%
	\providecommand \citenamefont [1]{#1}%
	\providecommand \href@noop [0]{\@secondoftwo}%
	\providecommand \href [0]{\begingroup \@sanitize@url \@href}%
	\providecommand \@href[1]{\@@startlink{#1}\@@href}%
	\providecommand \@@href[1]{\endgroup#1\@@endlink}%
	\providecommand \@sanitize@url [0]{\catcode `\\12\catcode `\$12\catcode
		`\&12\catcode `\#12\catcode `\^12\catcode `\_12\catcode `\%12\relax}%
	\providecommand \@@startlink[1]{}%
	\providecommand \@@endlink[0]{}%
	\providecommand \url  [0]{\begingroup\@sanitize@url \@url }%
	\providecommand \@url [1]{\endgroup\@href {#1}{\urlprefix }}%
	\providecommand \urlprefix  [0]{URL }%
	\providecommand \Eprint [0]{\href }%
	\providecommand \doibase [0]{https://doi.org/}%
	\providecommand \selectlanguage [0]{\@gobble}%
	\providecommand \bibinfo  [0]{\@secondoftwo}%
	\providecommand \bibfield  [0]{\@secondoftwo}%
	\providecommand \translation [1]{[#1]}%
	\providecommand \BibitemOpen [0]{}%
	\providecommand \bibitemStop [0]{}%
	\providecommand \bibitemNoStop [0]{.\EOS\space}%
	\providecommand \EOS [0]{\spacefactor3000\relax}%
	\providecommand \BibitemShut  [1]{\csname bibitem#1\endcsname}%
	\let\auto@bib@innerbib\@empty
	\bibitem [{\citenamefont {Koch}(1990)}]{koch1990diffusion}%
	\BibitemOpen
	\bibfield  {author} {\bibinfo {author} {\bibfnamefont {A.~L.}\ \bibnamefont
			{Koch}},\ }\bibfield  {title} {\bibinfo {title} {Diffusion the crucial
			process in many aspects of the biology of bacteria},\ }\href@noop {}
	{\bibfield  {journal} {\bibinfo  {journal} {Advances in microbial ecology}\
			,\ \bibinfo {pages} {37}} (\bibinfo {year} {1990})}\BibitemShut {NoStop}%
	\bibitem [{\citenamefont {Brangwynne}\ \emph {et~al.}(2009)\citenamefont
		{Brangwynne}, \citenamefont {Koenderink}, \citenamefont {MacKintosh},\ and\
		\citenamefont {Weitz}}]{brangwynne2009intracellular}%
	\BibitemOpen
	\bibfield  {author} {\bibinfo {author} {\bibfnamefont {C.~P.}\ \bibnamefont
			{Brangwynne}}, \bibinfo {author} {\bibfnamefont {G.~H.}\ \bibnamefont
			{Koenderink}}, \bibinfo {author} {\bibfnamefont {F.~C.}\ \bibnamefont
			{MacKintosh}},\ and\ \bibinfo {author} {\bibfnamefont {D.~A.}\ \bibnamefont
			{Weitz}},\ }\bibfield  {title} {\bibinfo {title} {Intracellular transport by
			active diffusion},\ }\href@noop {} {\bibfield  {journal} {\bibinfo  {journal}
			{Trends in cell biology}\ }\textbf {\bibinfo {volume} {19}},\ \bibinfo
		{pages} {423} (\bibinfo {year} {2009})}\BibitemShut {NoStop}%
	\bibitem [{\citenamefont {Soh}\ \emph {et~al.}(2010)\citenamefont {Soh},
		\citenamefont {Byrska}, \citenamefont {Kandere-Grzybowska},\ and\
		\citenamefont {Grzybowski}}]{soh2010reaction}%
	\BibitemOpen
	\bibfield  {author} {\bibinfo {author} {\bibfnamefont {S.}~\bibnamefont
			{Soh}}, \bibinfo {author} {\bibfnamefont {M.}~\bibnamefont {Byrska}},
		\bibinfo {author} {\bibfnamefont {K.}~\bibnamefont {Kandere-Grzybowska}},\
		and\ \bibinfo {author} {\bibfnamefont {B.~A.}\ \bibnamefont {Grzybowski}},\
	}\bibfield  {title} {\bibinfo {title} {Reaction-diffusion systems in
			intracellular molecular transport and control},\ }\href@noop {} {\bibfield
		{journal} {\bibinfo  {journal} {Angewandte Chemie International Edition}\
		}\textbf {\bibinfo {volume} {49}},\ \bibinfo {pages} {4170} (\bibinfo {year}
		{2010})}\BibitemShut {NoStop}%
	\bibitem [{\citenamefont {Wang}\ \emph {et~al.}(2010)\citenamefont {Wang},
		\citenamefont {Li},\ and\ \citenamefont {Pielak}}]{wang2010effects}%
	\BibitemOpen
	\bibfield  {author} {\bibinfo {author} {\bibfnamefont {Y.}~\bibnamefont
			{Wang}}, \bibinfo {author} {\bibfnamefont {C.}~\bibnamefont {Li}},\ and\
		\bibinfo {author} {\bibfnamefont {G.~J.}\ \bibnamefont {Pielak}},\ }\bibfield
	{title} {\bibinfo {title} {Effects of proteins on protein diffusion},\
	}\href@noop {} {\bibfield  {journal} {\bibinfo  {journal} {Journal of the
				American Chemical Society}\ }\textbf {\bibinfo {volume} {132}},\ \bibinfo
		{pages} {9392} (\bibinfo {year} {2010})}\BibitemShut {NoStop}%
	\bibitem [{\citenamefont {Mogre}\ \emph {et~al.}(2020)\citenamefont {Mogre},
		\citenamefont {Brown},\ and\ \citenamefont {Koslover}}]{mogre2020getting}%
	\BibitemOpen
	\bibfield  {author} {\bibinfo {author} {\bibfnamefont {S.~S.}\ \bibnamefont
			{Mogre}}, \bibinfo {author} {\bibfnamefont {A.~I.}\ \bibnamefont {Brown}},\
		and\ \bibinfo {author} {\bibfnamefont {E.~F.}\ \bibnamefont {Koslover}},\
	}\bibfield  {title} {\bibinfo {title} {Getting around the cell: physical
			transport in the intracellular world},\ }\href@noop {} {\bibfield  {journal}
		{\bibinfo  {journal} {Physical Biology}\ }\textbf {\bibinfo {volume} {17}},\
		\bibinfo {pages} {061003} (\bibinfo {year} {2020})}\BibitemShut {NoStop}%
	\bibitem [{\citenamefont {Dieteren}\ \emph {et~al.}(2011)\citenamefont
		{Dieteren}, \citenamefont {Gielen}, \citenamefont {Nijtmans}, \citenamefont
		{Smeitink}, \citenamefont {Swarts}, \citenamefont {Brock}, \citenamefont
		{Willems},\ and\ \citenamefont {Koopman}}]{dieteren2011solute}%
	\BibitemOpen
	\bibfield  {author} {\bibinfo {author} {\bibfnamefont {C.~E.}\ \bibnamefont
			{Dieteren}}, \bibinfo {author} {\bibfnamefont {S.~C.}\ \bibnamefont
			{Gielen}}, \bibinfo {author} {\bibfnamefont {L.~G.}\ \bibnamefont
			{Nijtmans}}, \bibinfo {author} {\bibfnamefont {J.~A.}\ \bibnamefont
			{Smeitink}}, \bibinfo {author} {\bibfnamefont {H.~G.}\ \bibnamefont
			{Swarts}}, \bibinfo {author} {\bibfnamefont {R.}~\bibnamefont {Brock}},
		\bibinfo {author} {\bibfnamefont {P.~H.}\ \bibnamefont {Willems}},\ and\
		\bibinfo {author} {\bibfnamefont {W.~J.}\ \bibnamefont {Koopman}},\
	}\bibfield  {title} {\bibinfo {title} {Solute diffusion is hindered in the
			mitochondrial matrix},\ }\href@noop {} {\bibfield  {journal} {\bibinfo
			{journal} {Proceedings of the National Academy of Sciences}\ }\textbf
		{\bibinfo {volume} {108}},\ \bibinfo {pages} {8657} (\bibinfo {year}
		{2011})}\BibitemShut {NoStop}%
	\bibitem [{\citenamefont {Means}\ \emph {et~al.}(2006)\citenamefont {Means},
		\citenamefont {Smith}, \citenamefont {Shepherd}, \citenamefont {Shadid},
		\citenamefont {Fowler}, \citenamefont {Wojcikiewicz}, \citenamefont {Mazel},
		\citenamefont {Smith},\ and\ \citenamefont {Wilson}}]{means2006reaction}%
	\BibitemOpen
	\bibfield  {author} {\bibinfo {author} {\bibfnamefont {S.}~\bibnamefont
			{Means}}, \bibinfo {author} {\bibfnamefont {A.~J.}\ \bibnamefont {Smith}},
		\bibinfo {author} {\bibfnamefont {J.}~\bibnamefont {Shepherd}}, \bibinfo
		{author} {\bibfnamefont {J.}~\bibnamefont {Shadid}}, \bibinfo {author}
		{\bibfnamefont {J.}~\bibnamefont {Fowler}}, \bibinfo {author} {\bibfnamefont
			{R.~J.}\ \bibnamefont {Wojcikiewicz}}, \bibinfo {author} {\bibfnamefont
			{T.}~\bibnamefont {Mazel}}, \bibinfo {author} {\bibfnamefont {G.~D.}\
			\bibnamefont {Smith}},\ and\ \bibinfo {author} {\bibfnamefont {B.~S.}\
			\bibnamefont {Wilson}},\ }\bibfield  {title} {\bibinfo {title} {Reaction
			diffusion modeling of calcium dynamics with realistic er geometry},\
	}\href@noop {} {\bibfield  {journal} {\bibinfo  {journal} {Biophysical
				journal}\ }\textbf {\bibinfo {volume} {91}},\ \bibinfo {pages} {537}
		(\bibinfo {year} {2006})}\BibitemShut {NoStop}%
	\bibitem [{\citenamefont {Brown}\ and\ \citenamefont
		{Rutenberg}(2015)}]{brown2015cluster}%
	\BibitemOpen
	\bibfield  {author} {\bibinfo {author} {\bibfnamefont {A.~I.}\ \bibnamefont
			{Brown}}\ and\ \bibinfo {author} {\bibfnamefont {A.~D.}\ \bibnamefont
			{Rutenberg}},\ }\bibfield  {title} {\bibinfo {title} {Cluster coarsening on
			drops exhibits strong and sudden size-selectivity},\ }\href@noop {}
	{\bibfield  {journal} {\bibinfo  {journal} {Soft Matter}\ }\textbf {\bibinfo
			{volume} {11}},\ \bibinfo {pages} {3786} (\bibinfo {year}
		{2015})}\BibitemShut {NoStop}%
	\bibitem [{\citenamefont {Brown}\ and\ \citenamefont
		{Rutenberg}(2017)}]{brown2017model}%
	\BibitemOpen
	\bibfield  {author} {\bibinfo {author} {\bibfnamefont {A.~I.}\ \bibnamefont
			{Brown}}\ and\ \bibinfo {author} {\bibfnamefont {A.~D.}\ \bibnamefont
			{Rutenberg}},\ }\bibfield  {title} {\bibinfo {title} {A model of autophagy
			size selectivity by receptor clustering on peroxisomes},\ }\href@noop {}
	{\bibfield  {journal} {\bibinfo  {journal} {Frontiers in Physics}\ }\textbf
		{\bibinfo {volume} {5}},\ \bibinfo {pages} {14} (\bibinfo {year}
		{2017})}\BibitemShut {NoStop}%
	\bibitem [{\citenamefont {Berg}\ and\ \citenamefont
		{Purcell}(1977)}]{berg1977physics}%
	\BibitemOpen
	\bibfield  {author} {\bibinfo {author} {\bibfnamefont {H.~C.}\ \bibnamefont
			{Berg}}\ and\ \bibinfo {author} {\bibfnamefont {E.~M.}\ \bibnamefont
			{Purcell}},\ }\bibfield  {title} {\bibinfo {title} {Physics of
			chemoreception},\ }\href@noop {} {\bibfield  {journal} {\bibinfo  {journal}
			{Biophysical journal}\ }\textbf {\bibinfo {volume} {20}},\ \bibinfo {pages}
		{193} (\bibinfo {year} {1977})}\BibitemShut {NoStop}%
	\bibitem [{\citenamefont {Yang}\ and\ \citenamefont
		{Koslover}(2023)}]{yang2023diffusive}%
	\BibitemOpen
	\bibfield  {author} {\bibinfo {author} {\bibfnamefont {Z.}~\bibnamefont
			{Yang}}\ and\ \bibinfo {author} {\bibfnamefont {E.~F.}\ \bibnamefont
			{Koslover}},\ }\bibfield  {title} {\bibinfo {title} {Diffusive exit rates
			through pores in membrane-enclosed structures},\ }\href@noop {} {\bibfield
		{journal} {\bibinfo  {journal} {Physical Biology}\ }\textbf {\bibinfo
			{volume} {20}},\ \bibinfo {pages} {026001} (\bibinfo {year}
		{2023})}\BibitemShut {NoStop}%
	\bibitem [{\citenamefont {Licata}\ and\ \citenamefont
		{Grill}(2009)}]{licata2009first}%
	\BibitemOpen
	\bibfield  {author} {\bibinfo {author} {\bibfnamefont {N.~A.}\ \bibnamefont
			{Licata}}\ and\ \bibinfo {author} {\bibfnamefont {S.~W.}\ \bibnamefont
			{Grill}},\ }\bibfield  {title} {\bibinfo {title} {The first-passage problem
			for diffusion through a cylindrical pore with sticky walls},\ }\href@noop {}
	{\bibfield  {journal} {\bibinfo  {journal} {The European Physical Journal E}\
		}\textbf {\bibinfo {volume} {30}},\ \bibinfo {pages} {439} (\bibinfo {year}
		{2009})}\BibitemShut {NoStop}%
	\bibitem [{\citenamefont {Schwarz}\ and\ \citenamefont
		{Blower}(2016)}]{schwarz2016endoplasmic}%
	\BibitemOpen
	\bibfield  {author} {\bibinfo {author} {\bibfnamefont {D.~S.}\ \bibnamefont
			{Schwarz}}\ and\ \bibinfo {author} {\bibfnamefont {M.~D.}\ \bibnamefont
			{Blower}},\ }\bibfield  {title} {\bibinfo {title} {The endoplasmic reticulum:
			structure, function and response to cellular signaling},\ }\href@noop {}
	{\bibfield  {journal} {\bibinfo  {journal} {Cellular and molecular life
				sciences}\ }\textbf {\bibinfo {volume} {73}},\ \bibinfo {pages} {79}
		(\bibinfo {year} {2016})}\BibitemShut {NoStop}%
	\bibitem [{\citenamefont {Westrate}\ \emph {et~al.}(2015)\citenamefont
		{Westrate}, \citenamefont {Lee}, \citenamefont {Prinz},\ and\ \citenamefont
		{Voeltz}}]{westrate2015form}%
	\BibitemOpen
	\bibfield  {author} {\bibinfo {author} {\bibfnamefont {L.}~\bibnamefont
			{Westrate}}, \bibinfo {author} {\bibfnamefont {J.}~\bibnamefont {Lee}},
		\bibinfo {author} {\bibfnamefont {W.}~\bibnamefont {Prinz}},\ and\ \bibinfo
		{author} {\bibfnamefont {G.}~\bibnamefont {Voeltz}},\ }\bibfield  {title}
	{\bibinfo {title} {Form follows function: the importance of endoplasmic
			reticulum shape},\ }\href@noop {} {\bibfield  {journal} {\bibinfo  {journal}
			{Annual review of biochemistry}\ }\textbf {\bibinfo {volume} {84}},\ \bibinfo
		{pages} {791} (\bibinfo {year} {2015})}\BibitemShut {NoStop}%
	\bibitem [{\citenamefont {Nixon-Abell}\ \emph {et~al.}(2016)\citenamefont
		{Nixon-Abell}, \citenamefont {Obara}, \citenamefont {Weigel}, \citenamefont
		{Li}, \citenamefont {Legant}, \citenamefont {Xu}, \citenamefont {Pasolli},
		\citenamefont {Harvey}, \citenamefont {Hess}, \citenamefont {Betzig} \emph
		{et~al.}}]{nixon2016increased}%
	\BibitemOpen
	\bibfield  {author} {\bibinfo {author} {\bibfnamefont {J.}~\bibnamefont
			{Nixon-Abell}}, \bibinfo {author} {\bibfnamefont {C.~J.}\ \bibnamefont
			{Obara}}, \bibinfo {author} {\bibfnamefont {A.~V.}\ \bibnamefont {Weigel}},
		\bibinfo {author} {\bibfnamefont {D.}~\bibnamefont {Li}}, \bibinfo {author}
		{\bibfnamefont {W.~R.}\ \bibnamefont {Legant}}, \bibinfo {author}
		{\bibfnamefont {C.~S.}\ \bibnamefont {Xu}}, \bibinfo {author} {\bibfnamefont
			{H.~A.}\ \bibnamefont {Pasolli}}, \bibinfo {author} {\bibfnamefont
			{K.}~\bibnamefont {Harvey}}, \bibinfo {author} {\bibfnamefont {H.~F.}\
			\bibnamefont {Hess}}, \bibinfo {author} {\bibfnamefont {E.}~\bibnamefont
			{Betzig}}, \emph {et~al.},\ }\bibfield  {title} {\bibinfo {title} {Increased
			spatiotemporal resolution reveals highly dynamic dense tubular matrices in
			the peripheral er},\ }\href@noop {} {\bibfield  {journal} {\bibinfo
			{journal} {Science}\ }\textbf {\bibinfo {volume} {354}},\ \bibinfo {pages}
		{aaf3928} (\bibinfo {year} {2016})}\BibitemShut {NoStop}%
	\bibitem [{\citenamefont {West}\ \emph {et~al.}(2011)\citenamefont {West},
		\citenamefont {Zurek}, \citenamefont {Hoenger},\ and\ \citenamefont
		{Voeltz}}]{west20113d}%
	\BibitemOpen
	\bibfield  {author} {\bibinfo {author} {\bibfnamefont {M.}~\bibnamefont
			{West}}, \bibinfo {author} {\bibfnamefont {N.}~\bibnamefont {Zurek}},
		\bibinfo {author} {\bibfnamefont {A.}~\bibnamefont {Hoenger}},\ and\ \bibinfo
		{author} {\bibfnamefont {G.~K.}\ \bibnamefont {Voeltz}},\ }\bibfield  {title}
	{\bibinfo {title} {A {3D} analysis of yeast {ER} structure reveals how {ER}
			domains are organized by membrane curvature},\ }\href@noop {} {\bibfield
		{journal} {\bibinfo  {journal} {Journal of Cell Biology}\ }\textbf {\bibinfo
			{volume} {193}},\ \bibinfo {pages} {333} (\bibinfo {year}
		{2011})}\BibitemShut {NoStop}%
	\bibitem [{\citenamefont {Georgiades}\ \emph {et~al.}(2017)\citenamefont
		{Georgiades}, \citenamefont {Allan}, \citenamefont {Wright}, \citenamefont
		{Woodman}, \citenamefont {Udommai}, \citenamefont {Chung},\ and\
		\citenamefont {Waigh}}]{georgiades2017flexibility}%
	\BibitemOpen
	\bibfield  {author} {\bibinfo {author} {\bibfnamefont {P.}~\bibnamefont
			{Georgiades}}, \bibinfo {author} {\bibfnamefont {V.~J.}\ \bibnamefont
			{Allan}}, \bibinfo {author} {\bibfnamefont {G.~D.}\ \bibnamefont {Wright}},
		\bibinfo {author} {\bibfnamefont {P.~G.}\ \bibnamefont {Woodman}}, \bibinfo
		{author} {\bibfnamefont {P.}~\bibnamefont {Udommai}}, \bibinfo {author}
		{\bibfnamefont {M.~A.}\ \bibnamefont {Chung}},\ and\ \bibinfo {author}
		{\bibfnamefont {T.~A.}\ \bibnamefont {Waigh}},\ }\bibfield  {title} {\bibinfo
		{title} {The flexibility and dynamics of the tubules in the endoplasmic
			reticulum},\ }\href@noop {} {\bibfield  {journal} {\bibinfo  {journal}
			{Scientific reports}\ }\textbf {\bibinfo {volume} {7}},\ \bibinfo {pages} {1}
		(\bibinfo {year} {2017})}\BibitemShut {NoStop}%
	\bibitem [{\citenamefont {Schroeder}\ \emph {et~al.}(2019)\citenamefont
		{Schroeder}, \citenamefont {Barentine}, \citenamefont {Merta}, \citenamefont
		{Schweighofer}, \citenamefont {Zhang}, \citenamefont {Baddeley},
		\citenamefont {Bewersdorf},\ and\ \citenamefont
		{Bahmanyar}}]{schroeder2019dynamic}%
	\BibitemOpen
	\bibfield  {author} {\bibinfo {author} {\bibfnamefont {L.~K.}\ \bibnamefont
			{Schroeder}}, \bibinfo {author} {\bibfnamefont {A.~E.}\ \bibnamefont
			{Barentine}}, \bibinfo {author} {\bibfnamefont {H.}~\bibnamefont {Merta}},
		\bibinfo {author} {\bibfnamefont {S.}~\bibnamefont {Schweighofer}}, \bibinfo
		{author} {\bibfnamefont {Y.}~\bibnamefont {Zhang}}, \bibinfo {author}
		{\bibfnamefont {D.}~\bibnamefont {Baddeley}}, \bibinfo {author}
		{\bibfnamefont {J.}~\bibnamefont {Bewersdorf}},\ and\ \bibinfo {author}
		{\bibfnamefont {S.}~\bibnamefont {Bahmanyar}},\ }\bibfield  {title} {\bibinfo
		{title} {Dynamic nanoscale morphology of the {ER} surveyed by {STED}
			microscopy},\ }\href@noop {} {\bibfield  {journal} {\bibinfo  {journal}
			{Journal of Cell Biology}\ }\textbf {\bibinfo {volume} {218}},\ \bibinfo
		{pages} {83} (\bibinfo {year} {2019})}\BibitemShut {NoStop}%
	\bibitem [{\citenamefont {Schuck}\ \emph {et~al.}(2009)\citenamefont {Schuck},
		\citenamefont {Prinz}, \citenamefont {Thorn}, \citenamefont {Voss},\ and\
		\citenamefont {Walter}}]{schuck2009membrane}%
	\BibitemOpen
	\bibfield  {author} {\bibinfo {author} {\bibfnamefont {S.}~\bibnamefont
			{Schuck}}, \bibinfo {author} {\bibfnamefont {W.~A.}\ \bibnamefont {Prinz}},
		\bibinfo {author} {\bibfnamefont {K.~S.}\ \bibnamefont {Thorn}}, \bibinfo
		{author} {\bibfnamefont {C.}~\bibnamefont {Voss}},\ and\ \bibinfo {author}
		{\bibfnamefont {P.}~\bibnamefont {Walter}},\ }\bibfield  {title} {\bibinfo
		{title} {Membrane expansion alleviates endoplasmic reticulum stress
			independently of the unfolded protein response},\ }\href@noop {} {\bibfield
		{journal} {\bibinfo  {journal} {Journal of Cell Biology}\ }\textbf {\bibinfo
			{volume} {187}},\ \bibinfo {pages} {525} (\bibinfo {year}
		{2009})}\BibitemShut {NoStop}%
	\bibitem [{\citenamefont {Bertolotti}\ \emph {et~al.}(2000)\citenamefont
		{Bertolotti}, \citenamefont {Zhang}, \citenamefont {Hendershot},
		\citenamefont {Harding},\ and\ \citenamefont {Ron}}]{bertolotti2000dynamic}%
	\BibitemOpen
	\bibfield  {author} {\bibinfo {author} {\bibfnamefont {A.}~\bibnamefont
			{Bertolotti}}, \bibinfo {author} {\bibfnamefont {Y.}~\bibnamefont {Zhang}},
		\bibinfo {author} {\bibfnamefont {L.~M.}\ \bibnamefont {Hendershot}},
		\bibinfo {author} {\bibfnamefont {H.~P.}\ \bibnamefont {Harding}},\ and\
		\bibinfo {author} {\bibfnamefont {D.}~\bibnamefont {Ron}},\ }\bibfield
	{title} {\bibinfo {title} {Dynamic interaction of {BiP} and {ER} stress
			transducers in the unfolded-protein response},\ }\href@noop {} {\bibfield
		{journal} {\bibinfo  {journal} {Nature Cell Biology}\ }\textbf {\bibinfo
			{volume} {2}},\ \bibinfo {pages} {326} (\bibinfo {year} {2000})}\BibitemShut
	{NoStop}%
	\bibitem [{\citenamefont {Brown}\ and\ \citenamefont
		{Koslover}(2021)}]{brown2021design}%
	\BibitemOpen
	\bibfield  {author} {\bibinfo {author} {\bibfnamefont {A.~I.}\ \bibnamefont
			{Brown}}\ and\ \bibinfo {author} {\bibfnamefont {E.~F.}\ \bibnamefont
			{Koslover}},\ }\bibfield  {title} {\bibinfo {title} {Design principles for
			the glycoprotein quality control pathway},\ }\href@noop {} {\bibfield
		{journal} {\bibinfo  {journal} {PLoS Computational Biology}\ }\textbf
		{\bibinfo {volume} {17}},\ \bibinfo {pages} {e1008654} (\bibinfo {year}
		{2021})}\BibitemShut {NoStop}%
	\bibitem [{\citenamefont {Weigel}\ \emph {et~al.}(2021)\citenamefont {Weigel},
		\citenamefont {Chang}, \citenamefont {Shtengel}, \citenamefont {Xu},
		\citenamefont {Hoffman}, \citenamefont {Freeman}, \citenamefont {Iyer},
		\citenamefont {Aaron}, \citenamefont {Khuon}, \citenamefont {Bogovic} \emph
		{et~al.}}]{weigel2021er}%
	\BibitemOpen
	\bibfield  {author} {\bibinfo {author} {\bibfnamefont {A.~V.}\ \bibnamefont
			{Weigel}}, \bibinfo {author} {\bibfnamefont {C.-L.}\ \bibnamefont {Chang}},
		\bibinfo {author} {\bibfnamefont {G.}~\bibnamefont {Shtengel}}, \bibinfo
		{author} {\bibfnamefont {C.~S.}\ \bibnamefont {Xu}}, \bibinfo {author}
		{\bibfnamefont {D.~P.}\ \bibnamefont {Hoffman}}, \bibinfo {author}
		{\bibfnamefont {M.}~\bibnamefont {Freeman}}, \bibinfo {author} {\bibfnamefont
			{N.}~\bibnamefont {Iyer}}, \bibinfo {author} {\bibfnamefont {J.}~\bibnamefont
			{Aaron}}, \bibinfo {author} {\bibfnamefont {S.}~\bibnamefont {Khuon}},
		\bibinfo {author} {\bibfnamefont {J.}~\bibnamefont {Bogovic}}, \emph
		{et~al.},\ }\bibfield  {title} {\bibinfo {title} {{ER}-to-{Golgi} protein
			delivery through an interwoven, tubular network extending from {ER}},\
	}\href@noop {} {\bibfield  {journal} {\bibinfo  {journal} {Cell}\ }\textbf
		{\bibinfo {volume} {184}},\ \bibinfo {pages} {2412} (\bibinfo {year}
		{2021})}\BibitemShut {NoStop}%
	\bibitem [{\citenamefont {Obara}\ \emph {et~al.}(2023)\citenamefont {Obara},
		\citenamefont {Moore},\ and\ \citenamefont
		{Lippincott-Schwartz}}]{obara2023structural}%
	\BibitemOpen
	\bibfield  {author} {\bibinfo {author} {\bibfnamefont {C.~J.}\ \bibnamefont
			{Obara}}, \bibinfo {author} {\bibfnamefont {A.~S.}\ \bibnamefont {Moore}},\
		and\ \bibinfo {author} {\bibfnamefont {J.}~\bibnamefont
			{Lippincott-Schwartz}},\ }\bibfield  {title} {\bibinfo {title} {Structural
			diversity within the endoplasmic reticulum—from the microscale to the
			nanoscale},\ }\href@noop {} {\bibfield  {journal} {\bibinfo  {journal} {Cold
				Spring Harbor Perspectives in Biology}\ }\textbf {\bibinfo {volume} {15}},\
		\bibinfo {pages} {a041259} (\bibinfo {year} {2023})}\BibitemShut {NoStop}%
	\bibitem [{\citenamefont {Zwanzig}(1992)}]{zwanzig1992diffusion}%
	\BibitemOpen
	\bibfield  {author} {\bibinfo {author} {\bibfnamefont {R.}~\bibnamefont
			{Zwanzig}},\ }\bibfield  {title} {\bibinfo {title} {Diffusion past an entropy
			barrier},\ }\href@noop {} {\bibfield  {journal} {\bibinfo  {journal} {The
				Journal of Physical Chemistry}\ }\textbf {\bibinfo {volume} {96}},\ \bibinfo
		{pages} {3926} (\bibinfo {year} {1992})}\BibitemShut {NoStop}%
	\bibitem [{\citenamefont {Reguera}\ and\ \citenamefont
		{Rubi}(2001)}]{reguera2001kinetic}%
	\BibitemOpen
	\bibfield  {author} {\bibinfo {author} {\bibfnamefont {D.}~\bibnamefont
			{Reguera}}\ and\ \bibinfo {author} {\bibfnamefont {J.}~\bibnamefont {Rubi}},\
	}\bibfield  {title} {\bibinfo {title} {Kinetic equations for diffusion in the
			presence of entropic barriers},\ }\href@noop {} {\bibfield  {journal}
		{\bibinfo  {journal} {Physical Review E}\ }\textbf {\bibinfo {volume} {64}},\
		\bibinfo {pages} {061106} (\bibinfo {year} {2001})}\BibitemShut {NoStop}%
	\bibitem [{\citenamefont {Huber}\ and\ \citenamefont
		{Wilkinson}(2019)}]{huber2019terasaki}%
	\BibitemOpen
	\bibfield  {author} {\bibinfo {author} {\bibfnamefont {G.}~\bibnamefont
			{Huber}}\ and\ \bibinfo {author} {\bibfnamefont {M.}~\bibnamefont
			{Wilkinson}},\ }\bibfield  {title} {\bibinfo {title} {Terasaki spiral ramps
			and intracellular diffusion},\ }\href@noop {} {\bibfield  {journal} {\bibinfo
			{journal} {Physical Biology}\ }\textbf {\bibinfo {volume} {16}},\ \bibinfo
		{pages} {065002} (\bibinfo {year} {2019})}\BibitemShut {NoStop}%
	\bibitem [{\citenamefont {Holcman}\ and\ \citenamefont
		{Schuss}(2013)}]{holcman2013control}%
	\BibitemOpen
	\bibfield  {author} {\bibinfo {author} {\bibfnamefont {D.}~\bibnamefont
			{Holcman}}\ and\ \bibinfo {author} {\bibfnamefont {Z.}~\bibnamefont
			{Schuss}},\ }\bibfield  {title} {\bibinfo {title} {Control of flux by narrow
			passages and hidden targets in cellular biology},\ }\href@noop {} {\bibfield
		{journal} {\bibinfo  {journal} {Reports on Progress in Physics}\ }\textbf
		{\bibinfo {volume} {76}},\ \bibinfo {pages} {074601} (\bibinfo {year}
		{2013})}\BibitemShut {NoStop}%
	\bibitem [{\citenamefont {Brown}\ \emph {et~al.}(2020)\citenamefont {Brown},
		\citenamefont {Westrate},\ and\ \citenamefont {Koslover}}]{brown2020impact}%
	\BibitemOpen
	\bibfield  {author} {\bibinfo {author} {\bibfnamefont {A.~I.}\ \bibnamefont
			{Brown}}, \bibinfo {author} {\bibfnamefont {L.~M.}\ \bibnamefont
			{Westrate}},\ and\ \bibinfo {author} {\bibfnamefont {E.~F.}\ \bibnamefont
			{Koslover}},\ }\bibfield  {title} {\bibinfo {title} {Impact of global
			structure on diffusive exploration of organelle networks},\ }\href@noop {}
	{\bibfield  {journal} {\bibinfo  {journal} {Scientific Reports}\ }\textbf
		{\bibinfo {volume} {10}},\ \bibinfo {pages} {4984} (\bibinfo {year}
		{2020})}\BibitemShut {NoStop}%
	\bibitem [{\citenamefont {Scott}\ \emph {et~al.}(2021)\citenamefont {Scott},
		\citenamefont {Brown}, \citenamefont {Mogre}, \citenamefont {Westrate},\ and\
		\citenamefont {Koslover}}]{scott2021diffusive}%
	\BibitemOpen
	\bibfield  {author} {\bibinfo {author} {\bibfnamefont {Z.~C.}\ \bibnamefont
			{Scott}}, \bibinfo {author} {\bibfnamefont {A.~I.}\ \bibnamefont {Brown}},
		\bibinfo {author} {\bibfnamefont {S.~S.}\ \bibnamefont {Mogre}}, \bibinfo
		{author} {\bibfnamefont {L.~M.}\ \bibnamefont {Westrate}},\ and\ \bibinfo
		{author} {\bibfnamefont {E.~F.}\ \bibnamefont {Koslover}},\ }\bibfield
	{title} {\bibinfo {title} {Diffusive search and trajectories on tubular
			networks: a propagator approach},\ }\href@noop {} {\bibfield  {journal}
		{\bibinfo  {journal} {The European Physical Journal E}\ }\textbf {\bibinfo
			{volume} {44}},\ \bibinfo {pages} {80} (\bibinfo {year} {2021})}\BibitemShut
	{NoStop}%
	\bibitem [{\citenamefont {Elam}\ \emph {et~al.}(2022)\citenamefont {Elam},
		\citenamefont {Fr{\'\i}as}, \citenamefont {Zhang}, \citenamefont {Rodal},\
		and\ \citenamefont {Fai}}]{elam2022fast}%
	\BibitemOpen
	\bibfield  {author} {\bibinfo {author} {\bibfnamefont {L.}~\bibnamefont
			{Elam}}, \bibinfo {author} {\bibfnamefont {M.~Q.}\ \bibnamefont
			{Fr{\'\i}as}}, \bibinfo {author} {\bibfnamefont {Y.}~\bibnamefont {Zhang}},
		\bibinfo {author} {\bibfnamefont {A.~A.}\ \bibnamefont {Rodal}},\ and\
		\bibinfo {author} {\bibfnamefont {T.~G.}\ \bibnamefont {Fai}},\ }\bibfield
	{title} {\bibinfo {title} {Fast solver for diffusive transport times on
			dynamic intracellular networks},\ }\href@noop {} {\bibfield  {journal}
		{\bibinfo  {journal} {arXiv preprint arXiv:2207.07682}\ } (\bibinfo {year}
		{2022})}\BibitemShut {NoStop}%
	\bibitem [{\citenamefont {Scott}\ \emph {et~al.}(2023)\citenamefont {Scott},
		\citenamefont {Koning}, \citenamefont {Vanderwerp}, \citenamefont {Cohen},
		\citenamefont {Westrate},\ and\ \citenamefont
		{Koslover}}]{scott2023endoplasmic}%
	\BibitemOpen
	\bibfield  {author} {\bibinfo {author} {\bibfnamefont {Z.~C.}\ \bibnamefont
			{Scott}}, \bibinfo {author} {\bibfnamefont {K.}~\bibnamefont {Koning}},
		\bibinfo {author} {\bibfnamefont {M.}~\bibnamefont {Vanderwerp}}, \bibinfo
		{author} {\bibfnamefont {L.}~\bibnamefont {Cohen}}, \bibinfo {author}
		{\bibfnamefont {L.~M.}\ \bibnamefont {Westrate}},\ and\ \bibinfo {author}
		{\bibfnamefont {E.~F.}\ \bibnamefont {Koslover}},\ }\bibfield  {title}
	{\bibinfo {title} {Endoplasmic reticulum network heterogeneity guides
			diffusive transport and kinetics},\ }\href@noop {} {\bibfield  {journal}
		{\bibinfo  {journal} {Biophysical Journal}\ } (\bibinfo {year}
		{2023})}\BibitemShut {NoStop}%
	\bibitem [{\citenamefont {Condamin}\ \emph {et~al.}(2007)\citenamefont
		{Condamin}, \citenamefont {B{\'e}nichou}, \citenamefont {Tejedor},
		\citenamefont {Voituriez},\ and\ \citenamefont
		{Klafter}}]{condamin2007first}%
	\BibitemOpen
	\bibfield  {author} {\bibinfo {author} {\bibfnamefont {S.}~\bibnamefont
			{Condamin}}, \bibinfo {author} {\bibfnamefont {O.}~\bibnamefont
			{B{\'e}nichou}}, \bibinfo {author} {\bibfnamefont {V.}~\bibnamefont
			{Tejedor}}, \bibinfo {author} {\bibfnamefont {R.}~\bibnamefont {Voituriez}},\
		and\ \bibinfo {author} {\bibfnamefont {J.}~\bibnamefont {Klafter}},\
	}\bibfield  {title} {\bibinfo {title} {First-passage times in complex
			scale-invariant media},\ }\href@noop {} {\bibfield  {journal} {\bibinfo
			{journal} {Nature}\ }\textbf {\bibinfo {volume} {450}},\ \bibinfo {pages}
		{77} (\bibinfo {year} {2007})}\BibitemShut {NoStop}%
	\bibitem [{\citenamefont {Obara}\ \emph {et~al.}(2024)\citenamefont {Obara},
		\citenamefont {Nixon-Abell}, \citenamefont {Moore}, \citenamefont {Riccio},
		\citenamefont {Hoffman}, \citenamefont {Shtengel}, \citenamefont {Xu},
		\citenamefont {Schaefer}, \citenamefont {Pasolli}, \citenamefont {Masson}
		\emph {et~al.}}]{obara2024motion}%
	\BibitemOpen
	\bibfield  {author} {\bibinfo {author} {\bibfnamefont {C.~J.}\ \bibnamefont
			{Obara}}, \bibinfo {author} {\bibfnamefont {J.}~\bibnamefont {Nixon-Abell}},
		\bibinfo {author} {\bibfnamefont {A.~S.}\ \bibnamefont {Moore}}, \bibinfo
		{author} {\bibfnamefont {F.}~\bibnamefont {Riccio}}, \bibinfo {author}
		{\bibfnamefont {D.~P.}\ \bibnamefont {Hoffman}}, \bibinfo {author}
		{\bibfnamefont {G.}~\bibnamefont {Shtengel}}, \bibinfo {author}
		{\bibfnamefont {C.~S.}\ \bibnamefont {Xu}}, \bibinfo {author} {\bibfnamefont
			{K.}~\bibnamefont {Schaefer}}, \bibinfo {author} {\bibfnamefont {H.~A.}\
			\bibnamefont {Pasolli}}, \bibinfo {author} {\bibfnamefont {J.-B.}\
			\bibnamefont {Masson}}, \emph {et~al.},\ }\bibfield  {title} {\bibinfo
		{title} {Motion of vapb molecules reveals er--mitochondria contact site
			subdomains},\ }\href@noop {} {\bibfield  {journal} {\bibinfo  {journal}
			{Nature}\ ,\ \bibinfo {pages} {1}} (\bibinfo {year} {2024})}\BibitemShut
	{NoStop}%
	\bibitem [{\citenamefont {B{\'e}nichou}\ \emph {et~al.}(2010)\citenamefont
		{B{\'e}nichou}, \citenamefont {Chevalier}, \citenamefont {Klafter},
		\citenamefont {Meyer},\ and\ \citenamefont
		{Voituriez}}]{benichou2010geometry}%
	\BibitemOpen
	\bibfield  {author} {\bibinfo {author} {\bibfnamefont {O.}~\bibnamefont
			{B{\'e}nichou}}, \bibinfo {author} {\bibfnamefont {C.}~\bibnamefont
			{Chevalier}}, \bibinfo {author} {\bibfnamefont {J.}~\bibnamefont {Klafter}},
		\bibinfo {author} {\bibfnamefont {B.}~\bibnamefont {Meyer}},\ and\ \bibinfo
		{author} {\bibfnamefont {R.}~\bibnamefont {Voituriez}},\ }\bibfield  {title}
	{\bibinfo {title} {Geometry-controlled kinetics},\ }\href@noop {} {\bibfield
		{journal} {\bibinfo  {journal} {Nature chemistry}\ }\textbf {\bibinfo
			{volume} {2}},\ \bibinfo {pages} {472} (\bibinfo {year} {2010})}\BibitemShut
	{NoStop}%
	\bibitem [{\citenamefont {Erickson}(2009)}]{erickson2009size}%
	\BibitemOpen
	\bibfield  {author} {\bibinfo {author} {\bibfnamefont {H.~P.}\ \bibnamefont
			{Erickson}},\ }\bibfield  {title} {\bibinfo {title} {Size and shape of
			protein molecules at the nanometer level determined by sedimentation, gel
			filtration, and electron microscopy},\ }\href@noop {} {\bibfield  {journal}
		{\bibinfo  {journal} {Biological procedures online}\ }\textbf {\bibinfo
			{volume} {11}},\ \bibinfo {pages} {32} (\bibinfo {year} {2009})}\BibitemShut
	{NoStop}%
	\bibitem [{\citenamefont {Mej{\'\i}a-Alvarez}\ \emph
		{et~al.}(1999)\citenamefont {Mej{\'\i}a-Alvarez}, \citenamefont {Kettlun},
		\citenamefont {R{\'\i}os}, \citenamefont {Stern},\ and\ \citenamefont
		{Fill}}]{mejia1999unitary}%
	\BibitemOpen
	\bibfield  {author} {\bibinfo {author} {\bibfnamefont {R.}~\bibnamefont
			{Mej{\'\i}a-Alvarez}}, \bibinfo {author} {\bibfnamefont {C.}~\bibnamefont
			{Kettlun}}, \bibinfo {author} {\bibfnamefont {E.}~\bibnamefont {R{\'\i}os}},
		\bibinfo {author} {\bibfnamefont {M.}~\bibnamefont {Stern}},\ and\ \bibinfo
		{author} {\bibfnamefont {M.}~\bibnamefont {Fill}},\ }\bibfield  {title}
	{\bibinfo {title} {Unitary ca2+ current through cardiac ryanodine receptor
			channels under quasi-physiological ionic conditions},\ }\href@noop {}
	{\bibfield  {journal} {\bibinfo  {journal} {The Journal of general
				physiology}\ }\textbf {\bibinfo {volume} {113}},\ \bibinfo {pages} {177}
		(\bibinfo {year} {1999})}\BibitemShut {NoStop}%
	\bibitem [{\citenamefont {Tran}\ \emph {et~al.}(2021)\citenamefont {Tran},
		\citenamefont {Carter}, \citenamefont {De~Mazi{\`e}re}, \citenamefont
		{Ashkenazi}, \citenamefont {Klumperman}, \citenamefont {Walter},\ and\
		\citenamefont {Jensen}}]{tran2021stress}%
	\BibitemOpen
	\bibfield  {author} {\bibinfo {author} {\bibfnamefont {N.-H.}\ \bibnamefont
			{Tran}}, \bibinfo {author} {\bibfnamefont {S.~D.}\ \bibnamefont {Carter}},
		\bibinfo {author} {\bibfnamefont {A.}~\bibnamefont {De~Mazi{\`e}re}},
		\bibinfo {author} {\bibfnamefont {A.}~\bibnamefont {Ashkenazi}}, \bibinfo
		{author} {\bibfnamefont {J.}~\bibnamefont {Klumperman}}, \bibinfo {author}
		{\bibfnamefont {P.}~\bibnamefont {Walter}},\ and\ \bibinfo {author}
		{\bibfnamefont {G.~J.}\ \bibnamefont {Jensen}},\ }\bibfield  {title}
	{\bibinfo {title} {The stress-sensing domain of activated {IRE1}$\alpha$
			forms helical filaments in narrow {ER} membrane tubes},\ }\href@noop {}
	{\bibfield  {journal} {\bibinfo  {journal} {Science}\ }\textbf {\bibinfo
			{volume} {374}},\ \bibinfo {pages} {52} (\bibinfo {year} {2021})}\BibitemShut
	{NoStop}%
	\bibitem [{\citenamefont {Bakunts}\ \emph {et~al.}(2017)\citenamefont
		{Bakunts}, \citenamefont {Orsi}, \citenamefont {Vitale}, \citenamefont
		{Cattaneo}, \citenamefont {Lari}, \citenamefont {Tade}, \citenamefont
		{Sitia}, \citenamefont {Raimondi}, \citenamefont {Bachi},\ and\ \citenamefont
		{van Anken}}]{bakunts2017ratiometric}%
	\BibitemOpen
	\bibfield  {author} {\bibinfo {author} {\bibfnamefont {A.}~\bibnamefont
			{Bakunts}}, \bibinfo {author} {\bibfnamefont {A.}~\bibnamefont {Orsi}},
		\bibinfo {author} {\bibfnamefont {M.}~\bibnamefont {Vitale}}, \bibinfo
		{author} {\bibfnamefont {A.}~\bibnamefont {Cattaneo}}, \bibinfo {author}
		{\bibfnamefont {F.}~\bibnamefont {Lari}}, \bibinfo {author} {\bibfnamefont
			{L.}~\bibnamefont {Tade}}, \bibinfo {author} {\bibfnamefont {R.}~\bibnamefont
			{Sitia}}, \bibinfo {author} {\bibfnamefont {A.}~\bibnamefont {Raimondi}},
		\bibinfo {author} {\bibfnamefont {A.}~\bibnamefont {Bachi}},\ and\ \bibinfo
		{author} {\bibfnamefont {E.}~\bibnamefont {van Anken}},\ }\bibfield  {title}
	{\bibinfo {title} {Ratiometric sensing of bip-client versus bip levels by the
			unfolded protein response determines its signaling amplitude},\ }\href@noop
	{} {\bibfield  {journal} {\bibinfo  {journal} {Elife}\ }\textbf {\bibinfo
			{volume} {6}},\ \bibinfo {pages} {e27518} (\bibinfo {year}
		{2017})}\BibitemShut {NoStop}%
	\bibitem [{\citenamefont {Kischuck}\ and\ \citenamefont
		{Brown}(2023)}]{kischuck2023tube}%
	\BibitemOpen
	\bibfield  {author} {\bibinfo {author} {\bibfnamefont {L.~T.}\ \bibnamefont
			{Kischuck}}\ and\ \bibinfo {author} {\bibfnamefont {A.~I.}\ \bibnamefont
			{Brown}},\ }\bibfield  {title} {\bibinfo {title} {Tube geometry controls
			protein cluster conformation and stability on the endoplasmic reticulum
			surface},\ }\href@noop {} {\bibfield  {journal} {\bibinfo  {journal} {Soft
				Matter}\ }\textbf {\bibinfo {volume} {19}},\ \bibinfo {pages} {6771}
		(\bibinfo {year} {2023})}\BibitemShut {NoStop}%
	\bibitem [{\citenamefont {Stroberg}\ \emph {et~al.}(2018)\citenamefont
		{Stroberg}, \citenamefont {Aktin}, \citenamefont {Savir},\ and\ \citenamefont
		{Schnell}}]{stroberg2018design}%
	\BibitemOpen
	\bibfield  {author} {\bibinfo {author} {\bibfnamefont {W.}~\bibnamefont
			{Stroberg}}, \bibinfo {author} {\bibfnamefont {H.}~\bibnamefont {Aktin}},
		\bibinfo {author} {\bibfnamefont {Y.}~\bibnamefont {Savir}},\ and\ \bibinfo
		{author} {\bibfnamefont {S.}~\bibnamefont {Schnell}},\ }\bibfield  {title}
	{\bibinfo {title} {How to design an optimal sensor network for the unfolded
			protein response},\ }\href@noop {} {\bibfield  {journal} {\bibinfo  {journal}
			{Molecular Biology of the Cell}\ }\textbf {\bibinfo {volume} {29}},\ \bibinfo
		{pages} {3052} (\bibinfo {year} {2018})}\BibitemShut {NoStop}%
	\bibitem [{\citenamefont {Redner}(2001)}]{redner2001guide}%
	\BibitemOpen
	\bibfield  {author} {\bibinfo {author} {\bibfnamefont {S.}~\bibnamefont
			{Redner}},\ }\href@noop {} {\emph {\bibinfo {title} {A guide to first-passage
				processes}}}\ (\bibinfo  {publisher} {Cambridge university press},\ \bibinfo
	{year} {2001})\BibitemShut {NoStop}%
	\bibitem [{\citenamefont {Collins}\ and\ \citenamefont
		{Kimball}(1949)}]{collins1949diffusion}%
	\BibitemOpen
	\bibfield  {author} {\bibinfo {author} {\bibfnamefont {F.~C.}\ \bibnamefont
			{Collins}}\ and\ \bibinfo {author} {\bibfnamefont {G.~E.}\ \bibnamefont
			{Kimball}},\ }\bibfield  {title} {\bibinfo {title} {Diffusion-controlled
			reaction rates},\ }\href@noop {} {\bibfield  {journal} {\bibinfo  {journal}
			{Journal of colloid science}\ }\textbf {\bibinfo {volume} {4}},\ \bibinfo
		{pages} {425} (\bibinfo {year} {1949})}\BibitemShut {NoStop}%
	\bibitem [{\citenamefont {H{\"a}nggi}\ \emph {et~al.}(1990)\citenamefont
		{H{\"a}nggi}, \citenamefont {Talkner},\ and\ \citenamefont
		{Borkovec}}]{hanggi1990reaction}%
	\BibitemOpen
	\bibfield  {author} {\bibinfo {author} {\bibfnamefont {P.}~\bibnamefont
			{H{\"a}nggi}}, \bibinfo {author} {\bibfnamefont {P.}~\bibnamefont
			{Talkner}},\ and\ \bibinfo {author} {\bibfnamefont {M.}~\bibnamefont
			{Borkovec}},\ }\bibfield  {title} {\bibinfo {title} {Reaction-rate theory:
			fifty years after kramers},\ }\href@noop {} {\bibfield  {journal} {\bibinfo
			{journal} {Reviews of modern physics}\ }\textbf {\bibinfo {volume} {62}},\
		\bibinfo {pages} {251} (\bibinfo {year} {1990})}\BibitemShut {NoStop}%
	\bibitem [{\citenamefont {Berg}\ and\ \citenamefont {von
			Hippel}(1985)}]{berg1985diffusion}%
	\BibitemOpen
	\bibfield  {author} {\bibinfo {author} {\bibfnamefont {O.~G.}\ \bibnamefont
			{Berg}}\ and\ \bibinfo {author} {\bibfnamefont {P.~H.}\ \bibnamefont {von
				Hippel}},\ }\bibfield  {title} {\bibinfo {title} {Diffusion-controlled
			macromolecular interactions},\ }\href@noop {} {\bibfield  {journal} {\bibinfo
			{journal} {Annual review of biophysics and biophysical chemistry}\ }\textbf
		{\bibinfo {volume} {14}},\ \bibinfo {pages} {131} (\bibinfo {year}
		{1985})}\BibitemShut {NoStop}%
	\bibitem [{\citenamefont {Grebenkov}\ \emph {et~al.}(2018)\citenamefont
		{Grebenkov}, \citenamefont {Metzler},\ and\ \citenamefont
		{Oshanin}}]{grebenkov2018strong}%
	\BibitemOpen
	\bibfield  {author} {\bibinfo {author} {\bibfnamefont {D.~S.}\ \bibnamefont
			{Grebenkov}}, \bibinfo {author} {\bibfnamefont {R.}~\bibnamefont {Metzler}},\
		and\ \bibinfo {author} {\bibfnamefont {G.}~\bibnamefont {Oshanin}},\
	}\bibfield  {title} {\bibinfo {title} {Strong defocusing of molecular
			reaction times results from an interplay of geometry and reaction control},\
	}\href@noop {} {\bibfield  {journal} {\bibinfo  {journal} {Communications
				Chemistry}\ }\textbf {\bibinfo {volume} {1}},\ \bibinfo {pages} {96}
		(\bibinfo {year} {2018})}\BibitemShut {NoStop}%
	\bibitem [{\citenamefont {Grebenkov}\ and\ \citenamefont
		{Skvortsov}(2022)}]{grebenkov2022mean}%
	\BibitemOpen
	\bibfield  {author} {\bibinfo {author} {\bibfnamefont {D.~S.}\ \bibnamefont
			{Grebenkov}}\ and\ \bibinfo {author} {\bibfnamefont {A.~T.}\ \bibnamefont
			{Skvortsov}},\ }\bibfield  {title} {\bibinfo {title} {Mean first-passage time
			to a small absorbing target in three-dimensional elongated domains},\
	}\href@noop {} {\bibfield  {journal} {\bibinfo  {journal} {Physical Review
				E}\ }\textbf {\bibinfo {volume} {105}},\ \bibinfo {pages} {054107} (\bibinfo
		{year} {2022})}\BibitemShut {NoStop}%
\end{thebibliography}
\end{document}